\documentclass[fleqn,usenatbib]{mnras}

\usepackage{newtxtext,newtxmath}

\usepackage[T1]{fontenc}

\DeclareRobustCommand{\VAN}[3]{#2}
\let\VANthebibliography\thebibliography
\def\thebibliography{\DeclareRobustCommand{\VAN}[3]{##3}\VANthebibliography}

\usepackage{graphicx}	
\usepackage{amsmath}	

\newcommand{\st}{{\it St}}
\title[evolution of pebble-accreting planets in discs]{On the evolution of pebble-accreting planets in evolving protoplanetary discs}
\author[A. Pierens]{Arnaud Pierens $^{1}$\thanks{E-mail:arnaud.pierens@u-bordeaux.fr} \\
$^{1}$Laboratoire d'astrophysique de Bordeaux, Univ. Bordeaux, CNRS, B18N, all\'ee Geoffroy Saint-Hilaire, 33615 Pessac, France}

\pagerange{\pageref{firstpage}--\pageref{lastpage}} 

\def\LaTeX{L\kern-.36em\raise.3ex\hbox{a}\kern-.15em 
    T\kern-.1667em\lower.7ex\hbox{E}\kern-.125emX}
\begin{document}
\label{firstpage}
\maketitle
\begin{abstract}

We examine  the migration of luminous low-mass cores in laminar protoplanetary discs where accretion occurs mainly because of disc winds and where the planet luminosity is  generated by pebble accretion. Using 2D hydrodynamical simulations, we determine the eccentricities induced by thermal forces as a function of gas and pebble accretion rates, and also evaluate the importance of the torque exerted by the solid component relative to the gas torque. \\
For a gas accretion rate $\dot M= 2\times 10^{-8}$ $M_\odot/$yr and pebble flux $\dot M_{peb}=170$ $M_\oplus$/Myr, we find that embryo eccentricities attain values comparable to the disc aspect ratio.  The planet radial excursion in the disc, however, causes the torque exerted by inflowing pebbles to cancel on average and migration to transition from outward to inward.  This is found to arise because the magnitude of thermal torques decreases exponentially with increasing eccentricity, and we provide a fitting formula for the thermal torque attenuation as a function of eccentricity.\\
  As the disc evolves, the accretion luminosity  becomes at some point too small to make the core eccentricity grow such that the solid component can exert a  non-zero  torque on the planet. This  torque is positive and for gas accretion rates $\dot M \lesssim 5\times 10^{-9}$  $M_\odot/$yr and pebble fluxes  $\dot M_{peb} \lesssim 120$ $M_\oplus/$Myr, it is found to overcome the gas torque  exerted on cores with mass $m_p\lesssim$ $1M_\oplus$,  resulting in outward migration.\\

\end{abstract}

\section{Introduction}

In protoplanetary discs  growth of particles beyond the mm-cm scale is rendered difficult due to the bouncing (Zsom et al. 2010) and fragmentation (Blum \& Wurm 2008) barriers.  The streaming instability (hereafter SI; Youdin \& Goodman 2005; Johansen et al. 2009; Simon et al. 2016)  appears as a promising process  to bypass these growth barriers  and may enable  the direct formation of 100-km sized planetesimals.  In the context of the SI, particles with Stokes number (or dimensionless stopping time) $\st \sim 0.001-0.1$ directly concentrate into clumps or filaments under the action of gas drag and can eventually collapse to form planetesimals with sizes up to $\sim 10^3$ km. These planetesimals can subsequently grow very efficiently by capturing inward drifting pebbles (Johansen \& Lacerda 2010; Lambrechts \& Johansen 2012), namely solids with Stokes number $\st \sim 0.01-1$ that are marginally coupled to the gas.   Once a mass of $\sim 0.01$  $M_\oplus$ is reached, pebble accretion can become very efficient, depending on the local conditions in the disc, and it has been shown that in a protoplanetary disc similar to the Minimum Mass Solar Nebula (MMSN), 10 Earth mass planets can be formed at 5 AU in $\sim 10^4$ years (Lambrechts \& Johansen 2012).  \\

As an embryo grows by accreting pebbles, potential energy released by the accretion of solids heats the disc locally, which can significantly impact its orbital evolution in the disc (Benitez-Llambay et al. 2015). Accretion heating indeed makes streamlines outflowing from the Hill sphere form  two underdense lobes leading/following the planet and which exert a torque on the planet. Because the orbital velocity of gas is slightly sub-Keplerian, the lobe located behind the planet tends to be less dense than the one located ahead of the planet, resulting in a net positive torque, which has been referred to as the "heating torque". 

It has been shown that heating forces can not only lead to the outward migration of embryos but also play an important role on the evolution of  eccentricity (Eklund \& Masset 2017; Chrenko et al. 2017;  Fromenteau \& Masset 2019). In particular, when the luminosity $L$ arising from the accretion of solids exceeds a critical value, the embryo can undergo exponential growth of its eccentricity with an associated growth rate $\propto L$. Core eccentricity is generally found to saturate to a value comparable to the disc aspect ratio. 

On the other hand, it has been shown by Liu \& Ormel (2018) that the pebble accretion efficiency increases with planet eccentricity, at least as long as the pebble-planet relative velocities remain small enough to enable for efficient pebble accretion. As noticed by Velasco Romero et al. (2022), this implies that there exists a feedback between the luminosity which determines the planet eccentricity through thermal forces, and the planet eccentricity which sets  the luminosity through the value of the pebble accretion efficiency. Interestingly, by examining the feedback of the eccentricity on the planet accretion rate, these authors obtained a self-consistent value of  eccentricity at saturation, as a function of Stokes number and disc parameters. \\

In this paper, we estimate the importance of thermal forces on the migration of pebble-accreting cores as a function of disc age, or equivalently as a function of the accretion rate and pebble flux through the disc.   Motivated by the fact that standard viscous discs have difficulties to explain the  typical accretion rates that have been observed around young stars (Hartmann et al. 1998), we rather assume that accretion proceeds through disk winds.  Recent ALMA observations indeed constrained $\alpha$ viscous stress parameters of $10^{-4}-10^{-3}$, at least at large distances $>20$ AU (Pinte et al. 2016; Flaherty et al. 2018), which is clearly too small to account for an accretion rate of  $\sim 10^{-8}$ $M_\odot$/yr .  Most sophisticated models of protoplanetary discs also suggest that the ionization degree outside $0.2$ AU might be too low to sustain MRI turbulence. In this region, it has been proposed that the disc remains essentially laminar and that accretion is mainly driven by a wind launched from high altitudes (Bai \& Stone 2013; Gressel et al. 2015; Bethune et al. 2017). A radial laminar flow  can also be induced in the disc midplane as a consequence of the Hall effect (Kunz 2008; Kunz \& Lesur 2013), and this is the situation that we consider in this paper.
Our main aim is to quantify how thermal forces  decrease as the disc evolves and the radial flux of pebbles drops, taking into account the feedback between luminosity and eccentricity and using a very simple model of pebble evolution. A related  issue that we address in the paper is the evolution of the torque exerted by the solid component onto the planet and its strength relatively to the gas torques.\\

This paper is organized as follows. In Sect. 2 and 3, we present the physical model and the numerical setup.  In Sect. 4.,  we describe the results of a fiducial simulation and discuss the effect of varying the accretion rate through the disc and the pebble flux in Sect. 5.  Finally,  we draw our conclusions in Sect. 6. 

\section{The physical model}
\label{sec:dustdisc}

  \subsection{Gas equations}
 We solve the hydrodynamical equations for the gas component in polar coordinates $(r,\varphi)$ (radial, azimuthal), with the origin of the frame located at the central star.   The governing equations  are the continuity equation which is given by:

\begin{equation}
\frac{\partial \Sigma_g}{\partial t}+\nabla\cdot(\Sigma_g{\mathbf v_g})=0,
\end{equation}

the Navier-Stokes equation which is given by:
\begin{equation}
\frac{\partial {\mathbf v_g}}{\partial t}+({\mathbf v_g}\cdot \nabla){\mathbf v_g}=-\frac{\nabla P}{\Sigma}-\frac{\Sigma_d}{\Sigma_g}\mathbf F_{drag}-{\bf \nabla} \Phi-\frac{\nabla\cdot \cal T}{\Sigma_g}+\frac{\gamma_w}{r}\mathbf e_\varphi,
\label{eq:gas}
\end{equation}

and the energy equation which is described in Sect. \ref{sec:energy}. In Eq. \ref{eq:gas},  $\Sigma_g$ and $\Sigma_d$ are the gas and pebble surface densities respectively, $\mathbf v_g$ the gas velocity, $\mathbf F_{drag}$ describes the frictional drag force  between the gas and the dust (see Sect.  \ref{sec:dustequations}), $P$ the gas pressure and $\cal T$ the viscous stress tensor (e.g. Nelson et al. 2000). To model viscous stresses arising from a small residual turbulent viscosity in the disc, we use a kinematic viscosity $\nu$ that is modelled using the $\alpha$ prescription (Shakura \& Sunyaev 1973) with $\alpha=10^{-4}$.

\subsubsection{Gravitational potential}

In Eq. \ref{eq:gas}, the gravitational potential $\Phi$  
includes the contributions from the star, planet and the indirect term. In this work, the gravitational influence of the planet on the gas disc is modelled using  a vertically averaged expression for the gravitational potential $\Phi_p$ (Muller \& Kley 2012, Chrenko et al. 2017): \\

\begin{equation}
\Phi_p=\frac{-Gm_p}{\Sigma_g}\int_{-z_{max}}^{z_{max}} \frac { \rho_g(z) dz }{\sqrt{|{\bf r}-{\bf r_p}|^2+r_s^2}}+\frac{Gm_p}{r_p^3}{\bf r}\cdot{\bf r_p}
\label{eq:phi_p}
\end{equation}

where $m_p$ is the planet mass, $r_s$ is a softening length which is set to $r_s=0.015H$ with $H$ the gas pressure scale height,  ${\bf r }$ is the vector pointing to the location in the disk  and  ${\bf r_p }$ is the vector pointing to the planet.   In the previous equation, the second term corresponds to the acceleration experienced by the centre of the reference frame due to the presence of the planet. The integral in Eq. \ref{eq:phi_p}  is computed by dividing the interval $[0,z_{max}]$ into $N_z=10$ equal intervals with $z_{max}=3H$,  and assuming hydrostatic equilibrium in the vertical direction such that the mass density of the gas $\rho_g(z)$ is given by:
\begin{equation}
\rho_g(z)=\frac{\Sigma_g}{\sqrt{2\pi}H}\exp\left(-\frac{z^2}{2H^2}\right)
\end{equation}

 The indirect term $\Phi_{ind}$ arising from the fact that the frame centred on the central star is not inertial  is given by (e.g. Nelson et al. 2000):

\begin{equation}
\Phi_{ind}=G\int_S\frac{(\Sigma+\Sigma_d)d{S'}}{R'^3}{\bf R}\cdot{\bf R'}
\end{equation}

where $S$ is the surface of the simulation domain.

\subsubsection{generating laminar accretion flows}

In this paper, we focus on advective discs for which accretion proceeds through a constant and inward laminar flow due to disc winds. We consider a scenario where the disc wind arises because of the Hall effect, and is induced by the Hall-Shear Instability (Kunz 2008;  Kunz \& Lesur 2013). In the case where the magnetic field and the disc angular momentum vector are aligned,  large radial and azimuthal fields can be generated, together with strong Maxwell stresses  throughout the vertical extent of the disc. Considering a 2D $(R,\phi)$ model of the flow would be reasonable simplification in that case,  as explained by McNally et al. (2017). Following Lega et al. (2022), to generate a laminar accretion with  constant accretion rate $\dot M_w$, we impose an  external, specific torque $\gamma_w$  with value:

\begin{equation}
\gamma_w=\frac{\dot M_w}{4\pi \Sigma_g}\sqrt{\frac{GM_\star}{R^3}}
\label{eq:gammaa}
\end{equation}

with $M_\star$ the mass of the central star. In practice, this can be done by adding a source term in the gas momentum equation corresponding to the last term in Eq. \ref{eq:gas}.

\subsubsection{Energy equation}
\label{sec:energy}

In addition to the continuity equation and momentum equation, we solve an energy equation that includes the effect of 
viscous heating, stellar irradiation, and radiative cooling. It reads:
\begin{equation}
\frac{\partial e}{\partial t}+\nabla \cdot (e{\bf v_g})=-(\gamma-1)e{\nabla \cdot {\bf v_g}}+Q^+_{vis}-Q^--2H\nabla  \cdot {\bf F}+Q_{acc}
\label{eq:energy}
\end{equation}
where $e$ is the thermal energy density,  $\gamma$ the adiabatic index which is set 
to $\gamma=1.4$. In the previous equation,  $Q^+_{vis}$ is the viscous heating term,  and $Q^-=2\sigma_B T_{eff}^4$ is the local radiative cooling from the disc surfaces, where 
$\sigma_B$ is the Stephan-Boltzmann constant and $T_{eff}$ the effective temperature which is given by (Menou \& Goodman 2004):
\begin{equation}
T_{eff}^4=\frac{T^4-T_{irr}^4}{\tau_{eff}} \quad \text{with} \quad \tau_{eff}=\frac{3}{8}\tau+\frac{\sqrt{3}}{4}+\frac{1}{4\tau}
\end{equation}
Here, $T$ is the midplane temperature and $\tau=\kappa\Sigma_g/2$ is the vertical optical depth, where  $\kappa$ is the Rosseland mean opacity which is taken from Bell \& Lin (1994). $T_{irr}$ is the irradiation temperature which is computed from the irradiation flux (Menou \& Goodman 2004):
\begin{equation}
\sigma_B T_{irr}^4= \frac{L_\star(1-\epsilon)}{4\pi R^2}\frac{H}{R}\left(\frac{d \log H}{d\log R}-1\right)
\end{equation}
where $\epsilon=1/2$ is the disc albedo, $L_\star$  is the stellar luminosity which is set to $L_\star=1.43 L_\odot$ , and where the factor $d \log H/d\log R$ is set to 
be $d \log H/d\log R=9/7$ (Chiang \& Goldreich 1997). This implies that self-shadowing effects are not taken into account in this study.\\
In Eq. \ref{eq:energy}, ${\bf F}$ is the radiative flux which is treated in the flux-limited  diffusion approach and which 
reads (e.g. Kley \& Crida 2008):
\begin{equation}
{\bf F}=-\frac{16\sigma_B \lambda T^3}{\rho_g \kappa}\nabla T
\end{equation}
 where $\lambda$ is a flux-limiter (e.g.  Kley 1989). Finally, $Q_{acc}=L_{acc}/S$ is the accretion heating term, with $S$ the cell area and $L_{acc}$ the luminosity of the accreting embryo (see Sect. \ref{sec:planets}).

  \subsection{Dust equations}
  \label{sec:dustequations}

The solid component is treated as a pressureless fluid whose equations for the conservation of mass and momentum are given by:

\begin{equation}
\frac{\partial \Sigma_d}{\partial t}+\nabla\cdot(\Sigma_d{\mathbf v_d})=-\dot \Sigma_d
\label{eq:dust_continuity}
\end{equation}

and

\begin{equation}
\frac{\partial {\mathbf v_d}}{\partial t}+({\mathbf v_d}\cdot \nabla){\mathbf v_d}=\mathbf F_{drag}-{\bf \nabla} \Phi -\frac{D_v}{t_s}\frac{\nabla \Sigma_d}{\Sigma_d}
\label{eq:dust}
\end{equation}

where $\mathbf v_d$ is the dust velocity, $t_s$ is the particle stopping time, $D_v$ is the turbulent (viscous) dust diffusion coefficient, and $\dot \Sigma_d$ represents the local dust density decrease resulting from pebble accretion (see Sec. \ref{sec:planets}).
 \subsubsection{Drag force between gas and dust}

In Eq. \ref{eq:dust}, the drag force resulting from the interaction with the gaseous disc is given by:
\begin{equation}
\mathbf F_{drag}=\frac{1}{t_s}({\mathbf v_g -\mathbf v_p})
\end{equation}
In the following, we parametrize the stopping time $t_s$  through the Stokes number $\st =t_s \Omega$ with $\Omega$ the Keplerian frequency. We  note that in the well-coupled regime, $t_s$ is related to the particle size $s_d$  through:
\begin{equation}
t_s=\frac{s_d\rho_d}{\Sigma_g \Omega}
\end{equation}

where $\rho_d$ is the particle internal density. Here, the size of dust grains is not fixed but we rather adopt a simple model of dust evolution  with a single representative size (Birnstiel et al. 2012). To this aim, we first calculate the maximum size of dust grains, assuming that it is limited by the effects of radial drift or fragmentation due to turbulence and differential drift.  The maximum size allowed by radial drift corresponds to a Stokes number   (e.g. Birnstiel et al. 2012; Kanagawa et al. 2018; Drazkowska et al. 2019):

\begin{equation}
\st_{\rm drift}=0.55\frac{1}{2|\eta|}\frac{\Sigma_d}{\Sigma_g}
\end{equation}

where $\eta=h^2(1-s-2f)$, with $h=H/R$ the disc aspect ratio, $s$  the negative power-law index of the gas surface density, and $f$ the flaring index.  The maximum size allowed by turbulence corresponds to a Stokes number   (e.g. Birnstiel et al. 2012; Kanagawa et al. 2018; Drazkowska et al. 2019):

\begin{equation}
\st_{\rm frag}=0.37\frac{v_f^2}{3\alpha c_s^2}
\end{equation}

where $v_f$ is the fragmentation velocity which is set to $v_f=10$ $m.s^{-1}$ and $c_s$ the sound speed. Fragmentation can also be induced by the effect of differential drift between grains which limits the maximum  Stokes number to a value of   (e.g. Birnstiel et al. 2012; Kanagawa et al. 2018; Drazkowska et al. 2019): 

\begin{equation}
\st_{\rm diff}=0.37\frac{v_f}{|\eta| v_k}
\end{equation}

Then, the representative size of dust grains is determined assuming that it corresponds to a Stokes number:

 \begin{equation}
\st=\min(\st_{\rm drift},\st_{\rm frag},\st_{\rm diff})
\label{eq:st}
\end{equation}

 \subsubsection{Dust diffusion}

 It has been shown by Tominaga et al. (2019) that introducing a diffusion term as a source term in the continuity equation for the dust can violate the conservation of the total angular momentum of the dusty  disc. To bypass this issue,  Klahr \& Schreiber (2020)  implemented the diffusion flux in the momentum equation for the dust rather than in the continuity equation. As a consequence, we model dust diffusion arising from the residual turbulence in the disc by including a diffusion pressure term within the dust momentum equation and corresponding to the last source term in Eq. \ref{eq:dust}. The corresponding dust diffusion coefficient is given by:

\begin{equation}
D_v=\frac{1+\st+4\st^2}{(1+\st^2)^2} \alpha c_s H
\label{eq:d}
\end{equation}

with $\alpha$ the viscous stress parameter.

 \subsection{Planets}
 \label{sec:planets}
  
We consider luminous planets that  initially evolve on fixed,  nearly circular orbits  with initial eccentricity $e_p=10^{-4}$,  and which can release heat in the disc as a consequence of  pebble accretion.  The   planet luminosity is related to the pebble accretion rate  onto the planet $\dot M_p$ through the relation:

\begin{equation}
L=\frac{Gm_p\dot M_p }{R_p}
\end{equation}

where $R_p$ is the physical radius of the planet which is calculated assuming a material density of $2$ $g.cm^{-3}$. Defining the pebble accretion efficiency as $\epsilon=\dot M_p/\dot M_{\rm peb}$ where  $\dot M_{\rm peb}$ is the inward mass flux of pebbles, it has been shown that $\epsilon$ depends mainly on $\eta$, $\st$, $\alpha$ and  on the planet eccentricity $e_p$.   Liu \& Ormel (2018) and Ormel \& Liu (2018) have derived prescriptions to calculate $\epsilon$ as a function of these parameters (see Appendix A) and that we employ to self-consistently determine the value of the planet luminosity. Thermal forces arising from heat release in the disc are indeed expected to make the planet eccentricity increase, and consequently impact the pebble accretion efficiency. As noted by Velasco Romero et al. (2021), there exists a feedback loop between the eccentricity growth driven by thermal forces and the luminosity  which tends to increase with eccentricity, at least in the regime where the eccentricity remains smaller than the disc aspect ratio. \\

Accretion of solids is accompanied by a reduction of the local gas surface density within the accretion radius of the planet  and which is represented  by the source term in the dust continuity equation (see Eq. \ref{eq:dust_continuity}). In practice, the solid density is reduced by a factor of $1-f_{\rm red} \Delta t $ at each timestep, where the reduction fator is $f_{\rm red}=\dot M_p/M_{\rm av}$ with $M_{\rm av}$ the amount of pebble mass available within the  accretion radius of the planet $R_{acc}$ whose value can be found in Appendix A. We find that a typical value for $f_{\rm red}$ corresponds to $f_{\rm red}\approx 0.5$ for the parameters that are adopted in the simulations. Also, we notice that for simplicity the removed mass is not added to the planet, such that we keep the planet mass constant.

\section{Numerical simulations}
\subsection{Numerical method}
Simulations were performed using the GENESIS (De Val-Borro et al. 2006) code which solves
the equations governing the disc evolution on a polar grid $(R,\varphi)$ using an advection scheme based on the monotonic  transport algorithm (Van Leer 1977). It uses the FARGO algorithm (Masset 2000) to avoid time step limitation due to the Keplerian velocity at the inner edge of the disc, and was recently extended to follow the evolution of a solid component that is modelled assuming a pressureless fluid.  Momentum exchange between the particles and the gas is handled by employing the semi-analytical scheme presented in Stoyanovskaya et al. (2018). This approach enables considering arbitrary solid concentrations and values for the Stokes number, and  is therefore very well suited for looking for solutions of non-stationary problems.  Tests of the numerical method to handle the momentum transfer between gas and dust have been presented in 
Pierens et al. (2019) and the code has been recently been used to study the non-linear evolution of the Secular Gravitational Instability (Pierens 2021).\\

The computational domain is covered by $N_R=2048$ radial grid cells uniformly distributed  between $R_{\rm in}=0.55$  and 
$R_{\rm out}=1.6$, and $N_\phi=4096$ azimuthal grid cells.  The computational units that we adopt are such that the unit of mass is the central mass $M_\star=1$ and is assumed to be equivalent to one Solar mass, the gravitational constant is $G=1$, and the distance  $R=1$ in the computational domain is set to  $5.2$ AU. In the following, this corresponds also to the initial semimajor axis $R_0$ of migrating planets so that when presenting the simulation results, time will be expressed in orbital periods at $R=R_0$.
 
 \subsection{Initial and boundary conditions}
 \label{sec:init}

We focus on disc models with  constant mass flow through the disc $\dot{M}=\dot M_v+\dot M_w$ where $\dot M_v=3\pi \alpha \Sigma_g H^2 \Omega$ is the accretion rate arising from viscous stresses, and $\dot M_w=2\pi R \Sigma_g v_{g,R}$ the  accretion rate due to the wind and for which we considered value  in the range $[2\times 10^{-9}, 2\times 10^{-8}]$ $M_\odot/$yr.  Due to the low value adopted for $\alpha$, 
we expect stellar heating to dominate over viscous heating, such that the  initial disc aspect ratio is
$h\propto (R/R_0)^{2/7}$ (Chiang \& Goldreich 1997). For a  disc with  constant $\dot M$, this implies an initial gas surface density   $\Sigma_g=\Sigma_0 (R/R_0)^{-15/14}$ 
(Bitsch et al. 2014) where $\Sigma_0$ is the surface density at $R=R_0$. For  $\dot M_w=2\times 10^{-8} M_\odot/$yr, $\Sigma_0$ is set to $\Sigma_0=6\times 10^{-4}$ which is equivalent to a MMSN disc with $\Sigma\approx 200$ $\rm{g}\cdot{\rm cm}^{-3}$ at 5 AU.  These profiles for the aspect ratio and surface density of the disc correspond to the ones that are adopted as initial conditions in our simulations. The initial radial velocity of the gas can then be constrained through the relation $v_{g,R}=\dot M_w/2\pi R \Sigma_g$. As the disc evolves in time and $\dot M$ progressively decreases, we assume that $v_r$ remains constant while $\Sigma_g$ is reduced according to the chosen value for $\dot M_w$. For instance for $\dot M_w=2\times 10^{-9} M_\odot/$yr, $\Sigma_0$ is  set to $\Sigma_0=6\times 10^{-5}$ in code units. We note that this assumes that the magnetic flux remains constant as the disc evolves, which might not be realistic (Lesur et al. 2022).    \\ 
At the outer boundary, gas enters the computational domain with a radial velocity $v_{g,R}$ while the surface density is fixed to its initial value. At the inner boundary, only the radial velocity is set to $v_{g,R}$ and the surface density is set to its value at $R_{\rm in}$.\\

\begin{figure}
\centering
\includegraphics[width=\columnwidth]{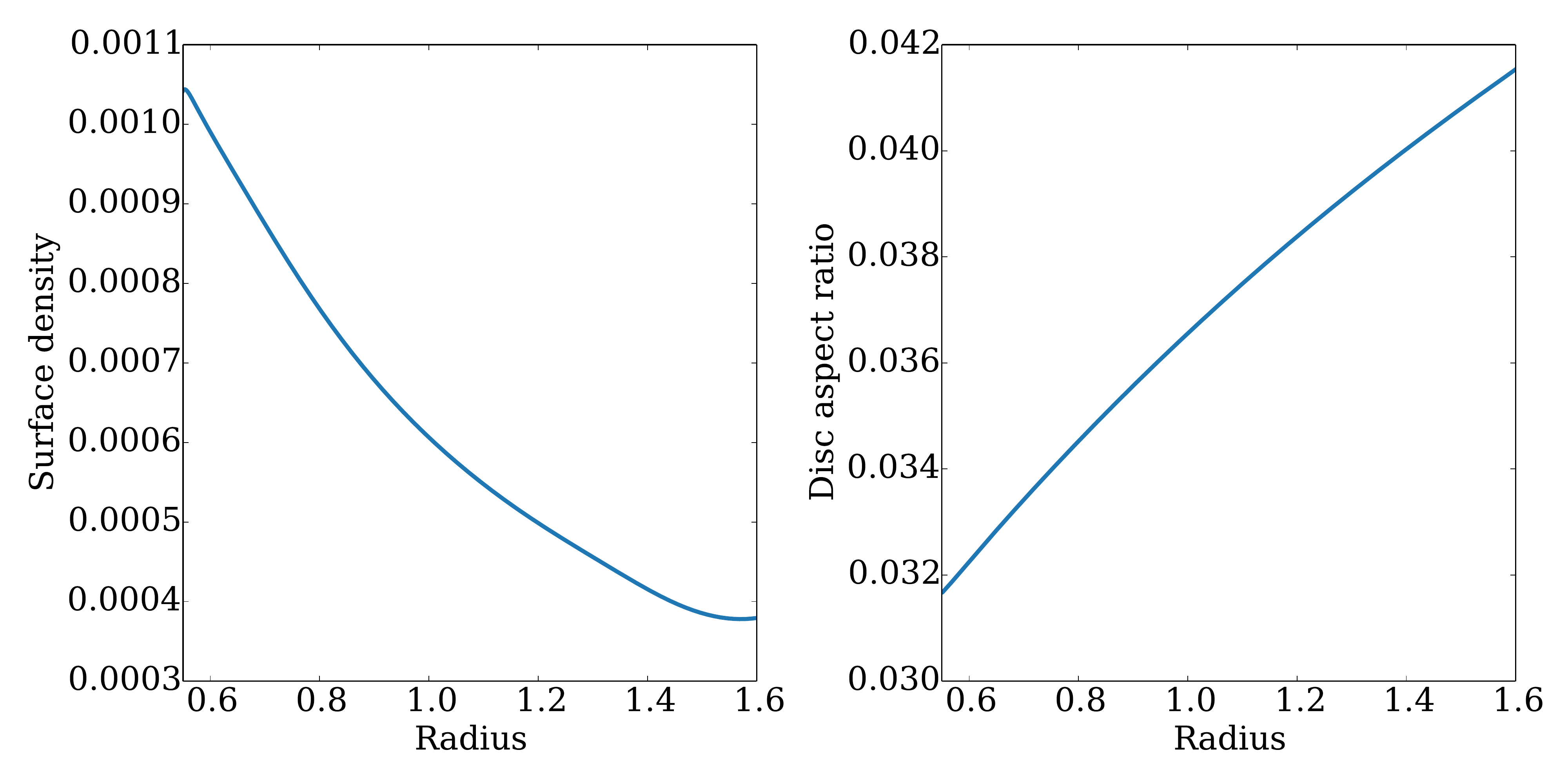}
\includegraphics[width=\columnwidth]{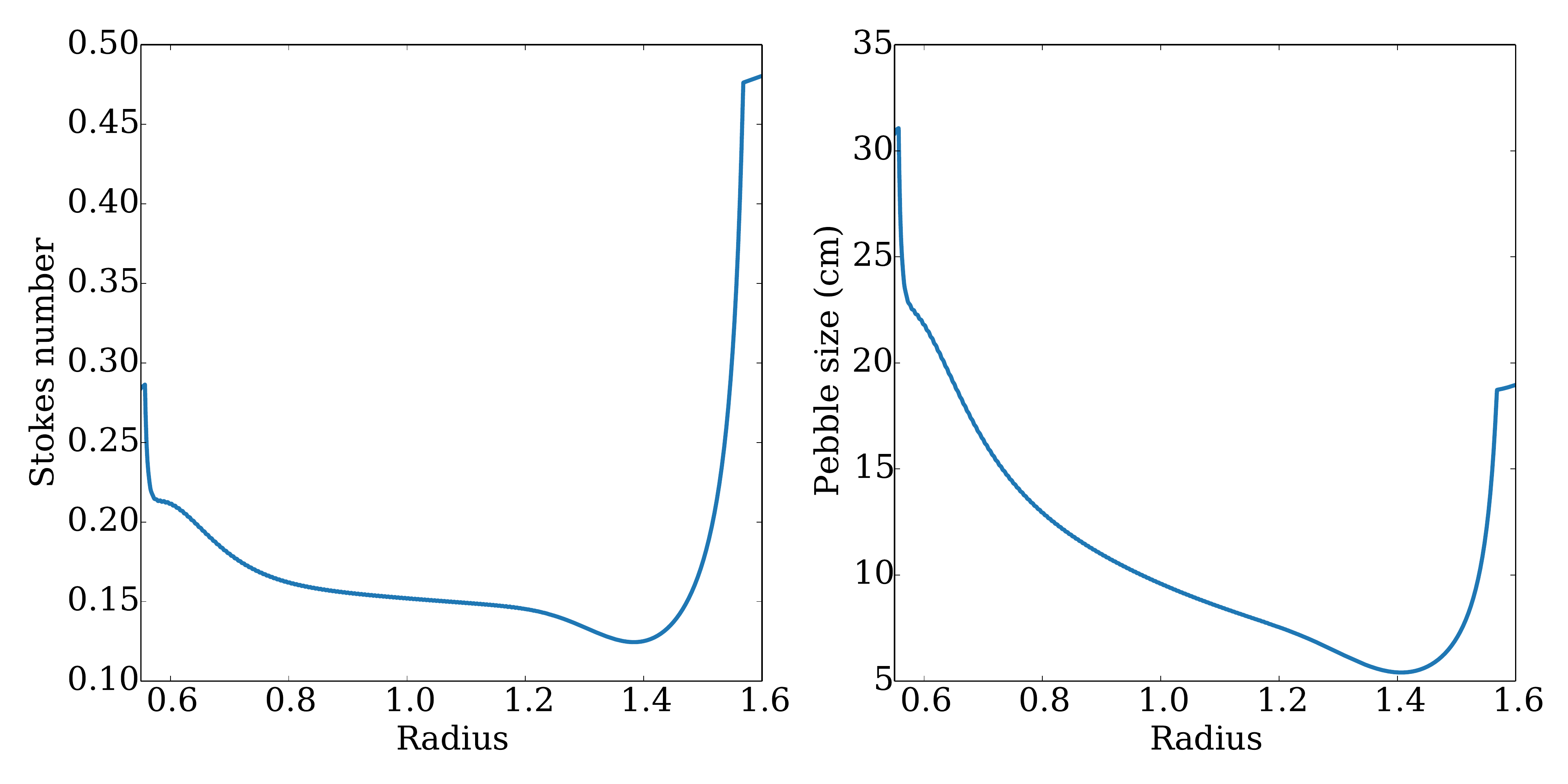}
\caption{Top panel: For $\dot M=2\times 10^{-8} M_\odot/$yr,  profiles of the gas surface density and aspect ratio at equilibrium.  Bottom panel: profiles of the Stokes number (left) and corresponding sizes (right) once a constant pebble flux through the disc has been reached. }
\label{fig:initial_profiles}
\end{figure}

Our simulations are split into two steps. We first evolve the gas and pebble discs until a quasi-stationary state is achieved, namely until the gas  and solid mass fluxes become constant through the disc, and then use the relaxed hydrodynamical quantities as initial conditions for runs with embedded planets.  In that case, we make use of wave-killing zones in the intervals $R\in [0.55,0.6]$ and $R\in [1.55,1.6]$ where each  variable f  is relaxed towards its initial value $f_0$ using the prescription of  de Val-Borro et al. (2006):

\begin{equation}
\frac{df}{dt}=-\frac{f-f_0}{\tau}{\cal P}(R)
\end{equation}

where   $\tau$ is the damping time-scale which is set to $0.1$ of the orbital period, and ${\cal P}(R)$ is a parabolic function which varies  between $1$ and $0$ from the edge of the domain to the inner edge of the damping zone.

The upper panel of Fig. \ref{fig:initial_profiles} shows the profiles of the gas surface density and aspect ratio at equilibrium once a constant accretion flow through the disc has been reached. At $R=1$, the aspect ratio is $h\approx 0.036$ and the gas surface density is $\Sigma_g\approx 6\times 10^{-4}$ so that the accretion rate due to viscous stress is estimated to be $\dot M_v\approx 7.3\times 10^{-10}$  in code units or $\dot M_v\approx 1.3\times 10^{-9} M_\odot/$yr when converting into physical units. Thus, the viscous accretion rate is $\approx 5\%$ of the accretion rate due to the wind,  independently of the value adopted for $\dot M_w$. \\

Observations indicate that the link between the disc age and mass accretion rate is given by (Hartmann et al. 1998; Bitsch et al. 2015):
\begin{equation}
\log\left(\frac{\dot M}{M_\odot/\rm yr}\right)=-8-1.4\log\left(\frac{t_{\rm disk}+10^5 {\rm yr}}{10^6 {\rm yr}}\right)
\label{eq:mdot}
\end{equation}
For a given value of $\dot M$, we follow Baumann \& Bitsch (2020) and set the pebble flux through the disc to:

\begin{equation}
\dot M_{\rm peb}=2\cdot 10^{-4}\exp\left(-\frac{t_{\rm disk}}{t_f}\right) \quad M_\oplus/{\rm yr}
\label{eq:mdot_peb}
\end{equation}

 where $t_{\rm disk}$ can be calculated for a given value of $\dot M$ from Eq. \ref{eq:mdot} and with  $t_f=3$ Myr. A constant pebble flux through the disc is obtained by imposing at the outer boundary a velocity corresponding to an estimation of the pebble radial drift velocity $v_{d,R}\approx -2 \eta \st R \Omega$ (Chrenko et al. 2017), with $\st$ given by Eq. \ref{eq:st}, together with a surface density given by 
$\Sigma_d=\dot M_{\rm peb}/2\pi R|v_{d,R}|$. For $\dot M_w=2\times 10^{-8} M_\odot/$yr, we plot in the bottom panel of Fig. \ref{fig:initial_profiles}, both $\st$ and the corresponding pebble size  as a function of radius.

\section{A reference run}
\label{sec:fiducial}

\begin{figure}
\centering
\includegraphics[width=\columnwidth]{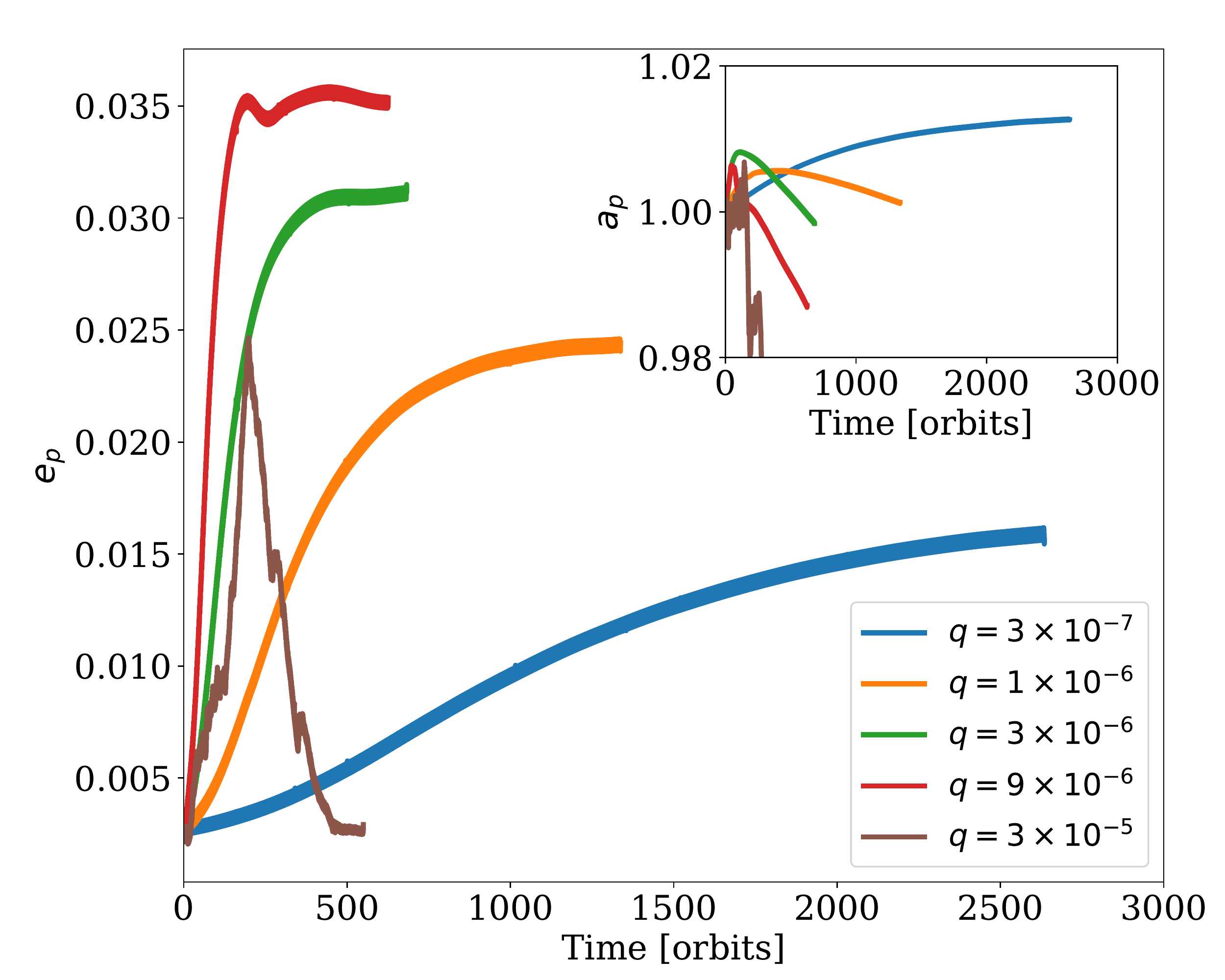}
\caption{For our reference run with $\dot M=2\times 10^{-8} M_\odot/$yr, evolution of the eccentricity and semimajor axis of luminous protoplanets. }
\label{fig:ae}
\end{figure}

In this section, we investigate the orbital evolution of proplanets in discs with $\dot M_w=2\times 10^{-8} M_\odot/$yr, and whose masses are in the range $m_p\in[0.1,10]$ $M_\oplus$, which corresponds to planet-to-star mass ratios  $q\in[3\times 10^{-7}, 3\times 10^{-5}]$.  The pebble flux is calculated according to Eq. \ref{eq:mdot_peb}, which gives $\dot M_{\rm peb}=170$ $M_\oplus/$Myr.  These planets are initiallly on circular orbits with semimajor axis $R_0=1$. 
For such low-mass planets, the half-width of the horseshoe region is estimated to be (Paardekooper et al. 2011):
\begin{equation}
x_s\approx 1.2 a_p\sqrt{q/h}
\end{equation}
with $a_p$ the planet semimajor axis. For the lowest mass that we consider $m_p=0.1$ $M_\oplus$ ($q=3\times 10^{-7}$), so that $x_s$ is resolved by approximately $7$ radial grid cells, which is enough to capture the dynamics of the horseshoe region with reasonable accuracy. Another scale that needs to be resolved with high accuracy is the size of the thermal disturbance $\lambda_c$ in the vicinity of the planet, and which is given by (Masset 2017): 

\begin{equation}
\lambda_c=\sqrt{\frac{\chi}{(3/2)\Omega \gamma}}
\end{equation}

where $\chi$ is the thermal diffusivity. Here, the thermal diffusivity is evaluated to $\chi \sim 1.3\times 10^{-5}$ at the initial planet location in dimensionless units, which results in $\lambda_c\sim 0.07H$. This lengthscale is therefore resolved by $\sim 5$ grid cell in the radial direction, and by $\sim 2$ grid cells in the azimuthal direction. This resolution is slightly smaller than the one that is required to obtain a reliable value of the planet eccentricity induced by thermal torques (Velasco Romero et al. 2021), so when presenting the results of the simulations in the following, we expect the asymptotic value reached by the planet eccentricity to be slightly underestimated.

\begin{figure}
\centering
\includegraphics[width=\columnwidth]{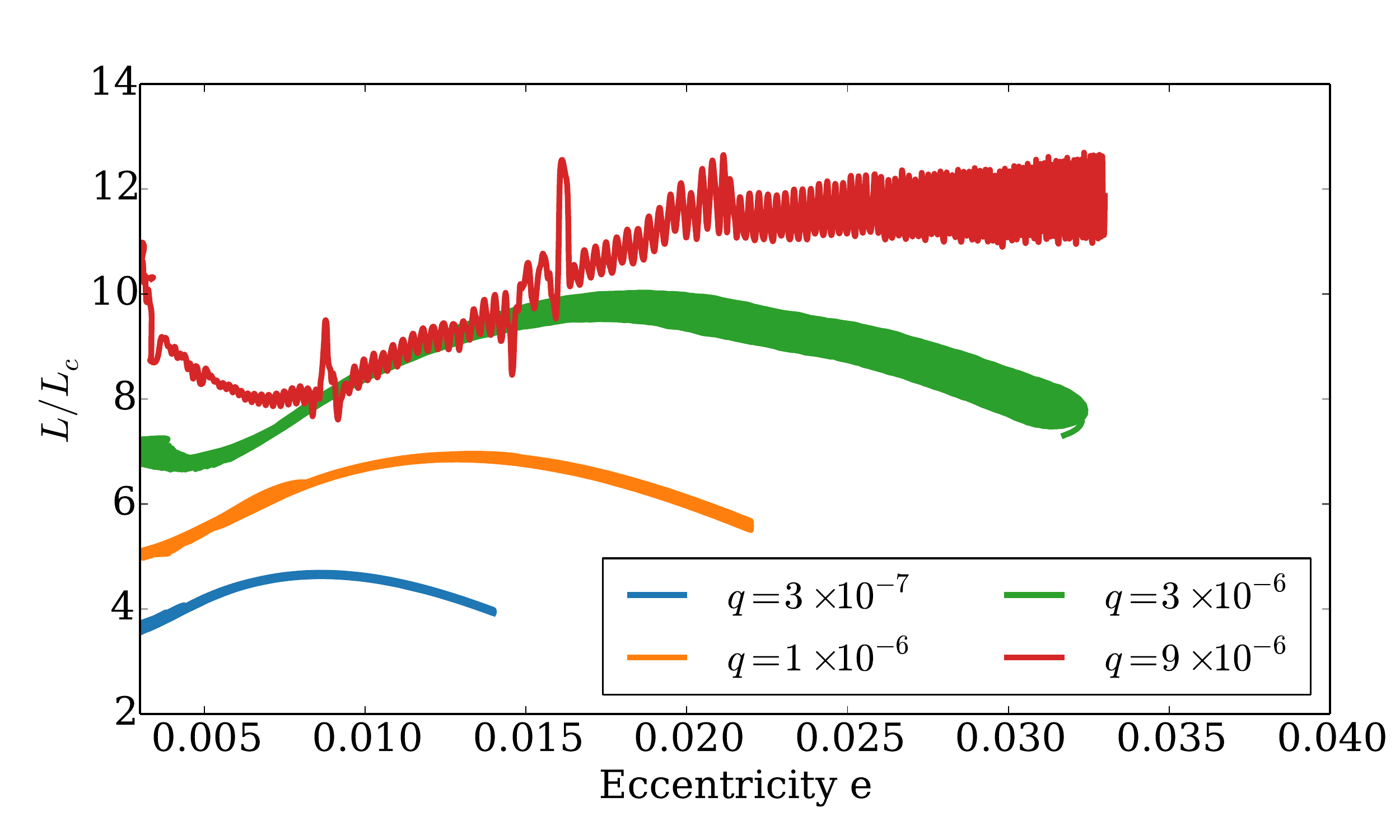}
\caption{For our reference run with $\dot M_w=2\times 10^{-8} M_\odot/$yr, embryo's luminosity as a function of eccentricity for planetary masses below the pebble isolation mass.  }
\label{fig:lumin}
\end{figure}

\subsection{Eccentricity evolution of protoplanets}

For our fiducial simulation, the time evolution of the eccentricity and semimajor axis is displayed in Fig. \ref{fig:ae}.  Initial orbital evolution involves growth of the planet eccentricity and outward migration as a consequence of the high pebble flux  inducing a strong heating torque. Planet eccentricities then saturate, except for the most massive planet that we consider ($q=3\times 10^{-5}$) whose eccentricity is quickly damped. This simply occurs because its mass is larger than the pebble isolation mass (Bitsch et al. 2018), causing the heating torque to cancel as the flux of pebbles at the planet location is stopped. For planetary masses below the pebble isolation mass, eccentricity growth is expected provided that the embryo's luminosity $L$ is higher than the critical luminosity $L_c$ given by (Masset 2017): 

\begin{equation}
L_c=\frac{4 \pi G m_p \chi \rho_g}{\gamma}
\end{equation}

We plot $L/L_c$ as a function of eccentricity in Fig. \ref{fig:lumin},  where we see that the embryo's luminosity tends indeed to be larger than the critical luminosity, giving rise to the period of eccentricity growth that is observed at early times. As noticed by Velasco Romero et al. (2021) the luminosity first increases with eccentricity due to the feedback loop that exists between these two quantities. Pebble accretion efficiency tends indeed to increase with the eccentricity $e_p$ (Liu \& Ormel 2018), resulting in a more luminous planet. In turn, an increase in the planet luminosity gives rise to a stronger thermal force, which promotes eccentricity growth. This is true provided that the eccentricity is not too large, otherwise pebble-planet encounters  may not satisfy the settling conditions for pebble accretion (Liu \& Ormel 2018). In our simulations,  we find that the luminosity starts to decrease as the eccentricity grows once the latter reaches $e_p\sim 0.01-0.02$, which is in good agreement with the results of Liu \& Ormel (2018).  Obviously, the maximum embryo's luminosity corresponds to the situation where the pebble accretion efficiency $\epsilon$ is maximum. For a $0.1$ $M_\oplus$ embryo, the maximum value for $\epsilon$ is found to be $\epsilon\sim 0.03$, whereas $\epsilon\sim 0.3$ for a $3$ $M_\oplus$ protoplanet. \\
Consistently with previous works (Eklund \& Masset 2017; Velasco Romero et al. 2021), planet eccentricities are found to saturate to  values comparable to the disc aspect ratio. The corresponding values of the planet luminosity can be obtained from Fig. \ref{fig:lumin}. We see that for a  $0.3$ $M_\oplus$ embryo ($q=10^{-6}$), $e_p\sim 0.025$ and $L\sim 5L_c$ at equilibrium, whereas  $e_p\sim 0.03$ and $L\sim 8L_c$ for a   $1$ $M_\oplus$ protoplanet.  These values can be directy compared to the results of  Velasco Romero et al. (2021) who calculated the equilibrium eccentricities and luminosities as a function of disc parameters, taking into account the feedback between eccentricity and luminosity.  Here, we have $\eta \approx 10^{-3}$, $\alpha=10^{-4}$ and $\st \sim 0.15$ (see Fig. \ref{fig:initial_profiles}) and for a similar set of parameters, Velasco Romero et al. (2021) found $e_p\sim 0.03$ for $m_p=0.3$ $M_\oplus$ and $e_p\sim 0.04$ for $m_p=1$ $M_\oplus$ (see Fig. 9 in Velasco Romero et al. 2021), which is close to the values that we report from our simulations. These authors found however slightly smaller equilibrium luminosities in comparison to our results, with $L\approx1-5$ $L_c$ for $m_p\in [0.3,3]$ $M_\oplus$ (see their Fig. 10).   

\subsection{Semimajor axis evolution of protoplanets }
\label{sec:a}

Turning back to Fig. \ref{fig:ae}, we see that after the initial outward migration stage, all but the lowest mass that we consider systematically migrate inward. A similar behaviour was observed by Eklund \& Masset (2017), who suggested this to be a consequence of the reduction of a positive corotation torque as the eccentricity increases (Fendyke \& Nelson 2012). The corotation torque is generally composed of a barotropic part which scales with the vortensity gradient  plus an entropy-related torque which scales with the entropy gradient (Paardekooper et al. 2011). Here, $\Sigma \propto R^{-15/14}$ so the vortensity component of the corotation torque is positive and the initial entropy profile is $S\propto R^{-0.05}$ such that the  entropy component of the corotation torque is also positive, although rather small. Focusing on the vortensity part of the corotation torque, its maximum value is reached when the viscous timescale across the horseshoe region $\tau_v=x_s^2/\nu$ is approximately equal to half the libration time-scale $\tau_{lib}=8\pi a_p/(3\Omega x_s)$, which corresponds to planet-to-star mass ratios $q\approx 1.7\times 10^{-6}$. Hence, for planets in this mass range migration reversal may at first sight be compatible with a cut-off of the corotation torque induced by the radial excursion of the planet. However, change in the migration direction is also observed for a planet with $q\approx 9\times 10^{-6}$,  and whose corotation torque is expected to be partly saturated. This suggests that an alternative process might be required to cause migration reversal in that case.  We will come back to this issue in more details in Sect. \ref{sec:torque}. \\

We remind the reader that a change in the  direction of migration does not necessarily mean a change in the sign of the torque exerted on the planet, since we have (e.g. Bitsch \& Kley 2010):

\begin{equation}
\frac{\Gamma}{L_p}=\frac{\dot a_p}{a_p}-\frac{e_p^2}{1-e_p^2}\frac{\dot e_p}{_p}
\end{equation}

where $\Gamma$ is the total torque and $L_p$ the planet angular momentum. The previous relation shows that the torque contributes also to the change in the planet  eccentricity. In the subsonic case, however, namely for $e_p<h$, we can be confident that migration reversal is triggered as soon as the total torque changes sign (Ida et al. 2020). Although not shown here, we checked by  looking at the evolution of the total torques and by close inspection of the semimajor axis evolution in  Fig. \ref{fig:ae} that this is indeed the case. 
\subsection{Torque evolution}
\label{sec:torque}

\begin{figure*}
\centering
\includegraphics[width=\columnwidth]{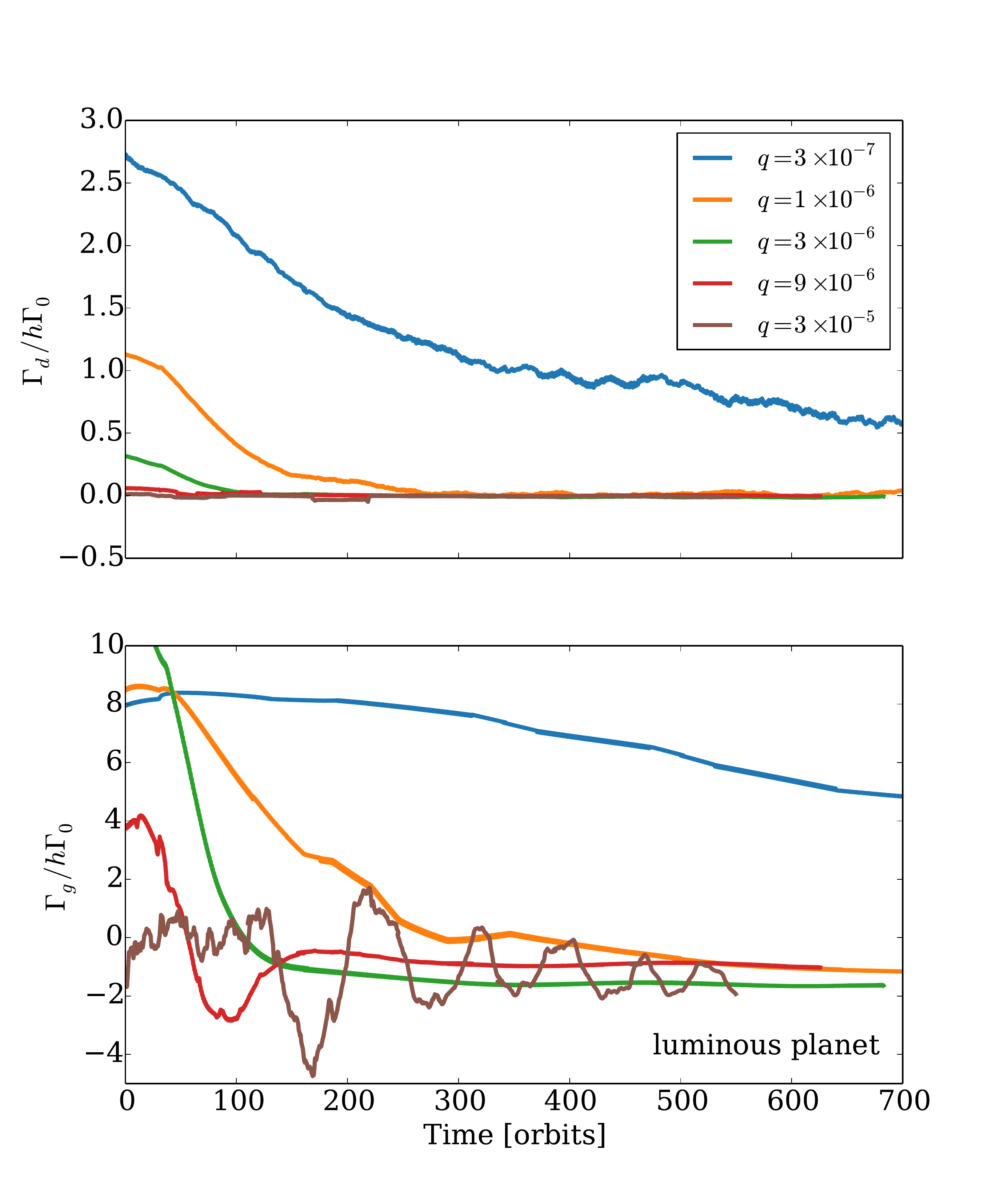}
\includegraphics[width=\columnwidth]{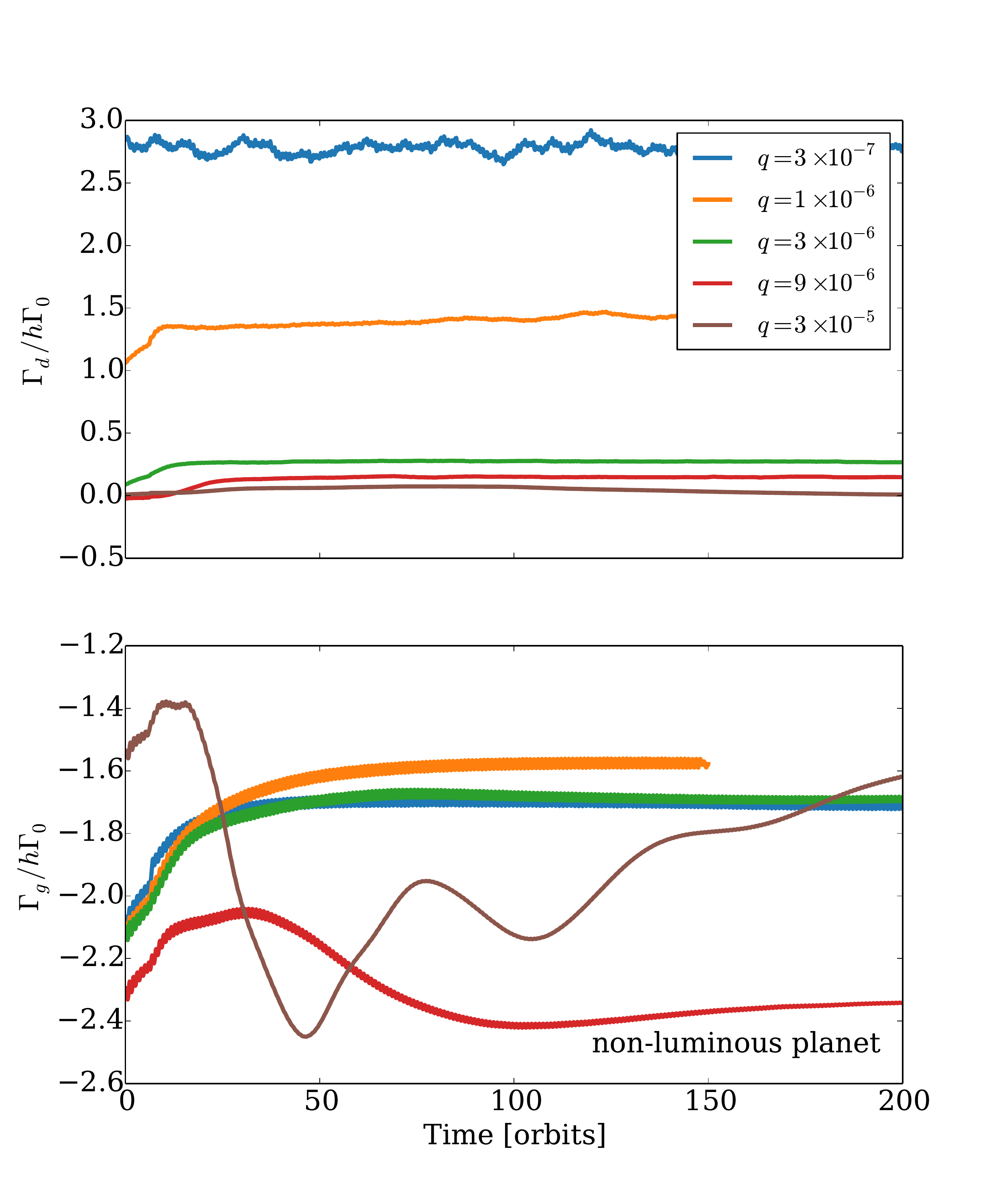}
\caption{For our reference run with $\dot M=2\times 10^{-8} M_\odot/$yr, evolution of the pebble and gas torques as a function of time. The left panel corresponds to the case of  a luminous planet whereas the right panel corresponds to the case of a non-luminous planet.}
\label{fig:torques}
\end{figure*}

In this work, the total torque acting on a planet is composed of the gas torque $\Gamma_g$ plus the torque exerted by incoming pebbles $\Gamma_d$. We show in the left panel of  Fig. \ref{fig:torques} the time evolution of $\Gamma_g/h\Gamma_0$ and $\Gamma_d/h\Gamma_0$ with:

\begin{equation}
\Gamma_0=\Sigma_p a_p^4 \Omega_p^2 q^2 h^{-3}
\end{equation}

and where $\Sigma_p$ is the gas surface density at the planet location and $\Omega_p$ the planet angular velocity. For comparison, we also plot these two components for a non-luminous planet in the right panel of  Fig. \ref{fig:torques}. Consistently with Regaly (2020), the solid torque tends to be positive when pebble accretion is accounted for, due to the formation of an underdense solid pattern located behind the planet (Regaly 2020). Interestingly, we see that the solid torque overcomes the gas torque for $q=3\times 10^{-7}$, such that migration is driven by pebbles in that case. 

Regarding the gas torque, the continuous increase in the torque that is observed for $q=3\times 10^{-5}$ is probably due to dynamical corotation torques operating (McNally et al. 2017, 2018). Using the same definitions  as in McNally et al. (2017), we define $\tau_f$ as the time for a fluid element to cross the corotation region due to the radial gas inflow:

$$
\tau_f=\frac{2x_s}{-v_{g,R}}
$$

and $\xi=\tau_f/\tau_{lib}$ as the parameter controlling the modification of the horseshoe region due the radial inflow of gas. Given that $-v_{g,R}\sim 10^{-5}$ in dimensionless units, we get $\xi \sim 30$ which confirms that a $10$ $M_\oplus$ planet can be subject to  a strong dynamical corotation torque (McNally et al. 2017). Moreover, the initial planet migration velocity is to estimated to $v_{p,R}\sim 3.5\times 10^{-5}$ in code units so that the evolution outcome in that case is expected to be inward migration with the planet migrating slightly faster than the gas inflow speed (McNally et al. 2018). For planets in the Earth-mass regime, however, the corotation torque is essentially unsaturated with the consequence that the dynamical corotation torque is negligible in that  case. \\

\begin{figure*}
\centering
\includegraphics[width=\textwidth]{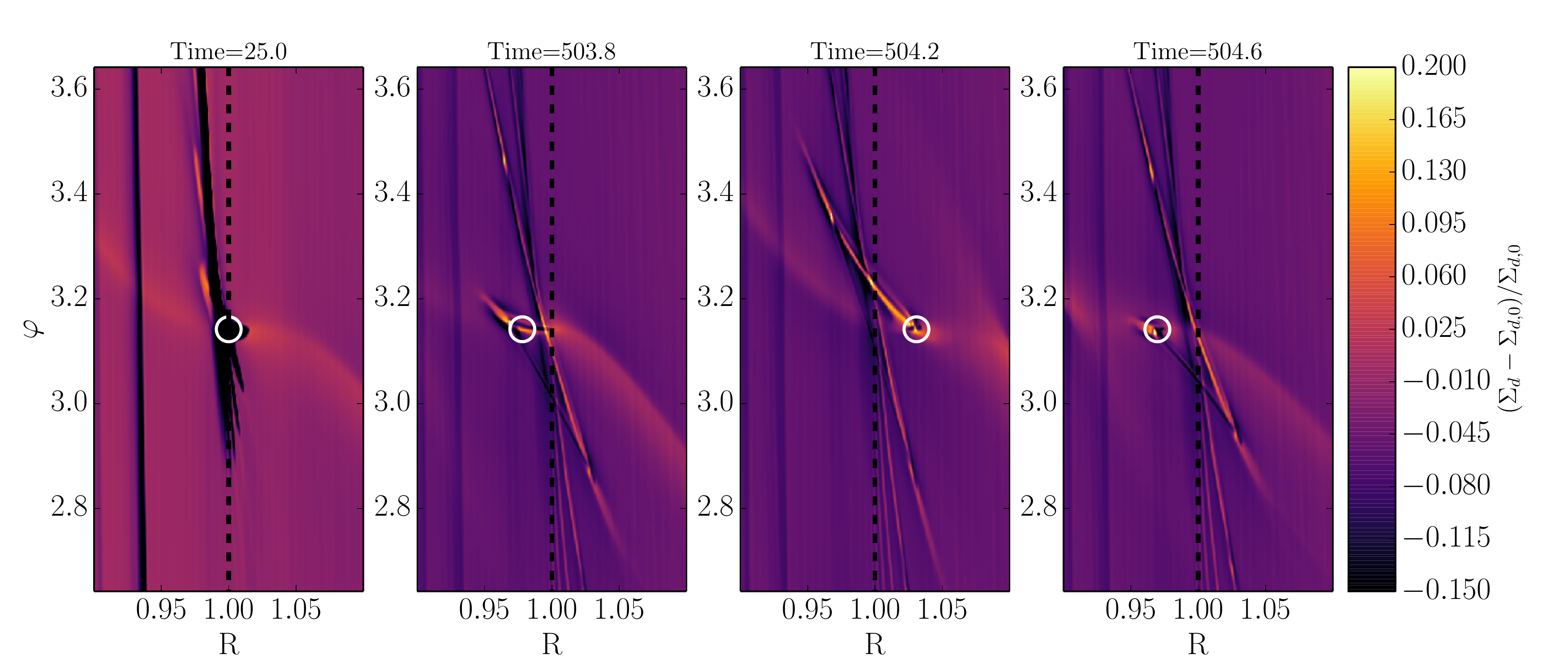}
\includegraphics[width=\textwidth]{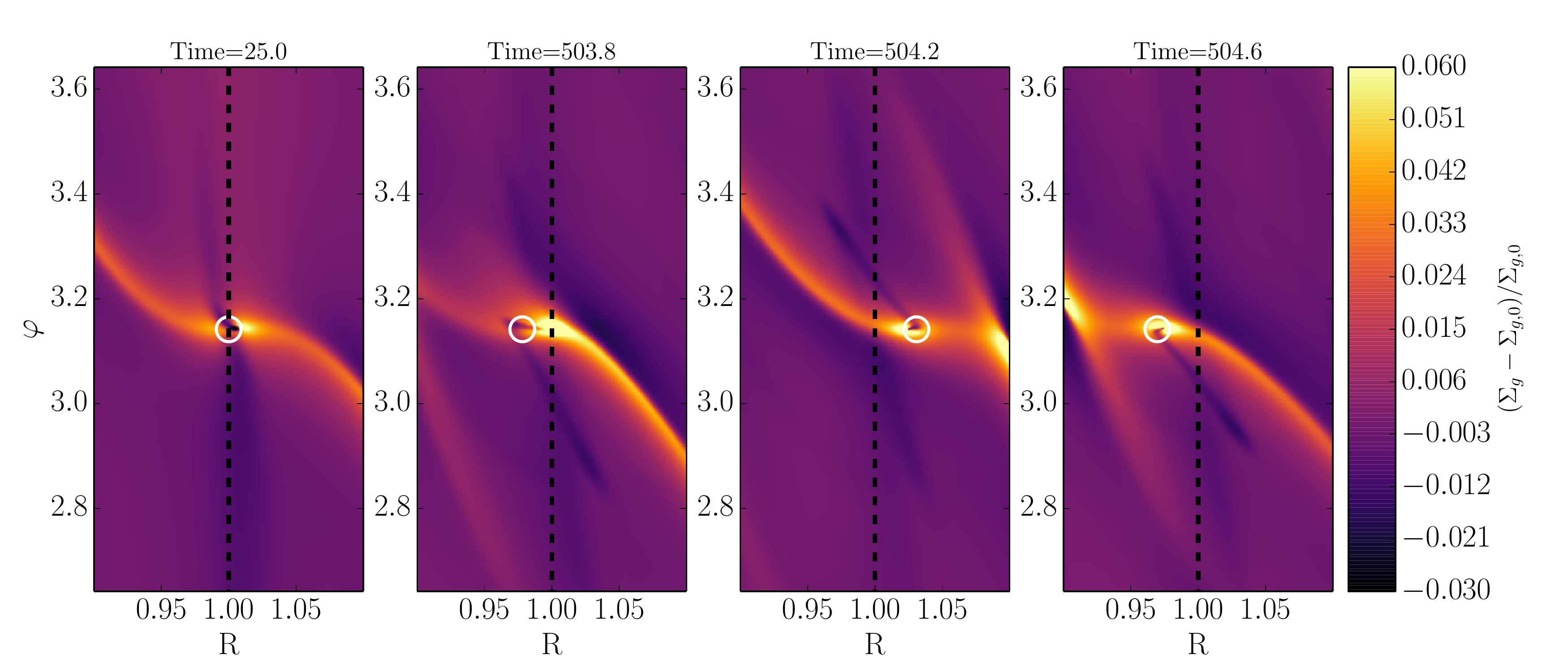}
\caption{For our reference run with $\dot M=2\times 10^{-8} M_\odot/$yr and a luminous $1$ $M_\oplus$ planet, relative surface density of the solid (top panel) and gas (bottom panel) components at different times. At times $t>500$, the planet eccentricity has reached a saturated value. }
\label{fig:2d}
\end{figure*}

\begin{figure}
\centering
\includegraphics[width=\columnwidth]{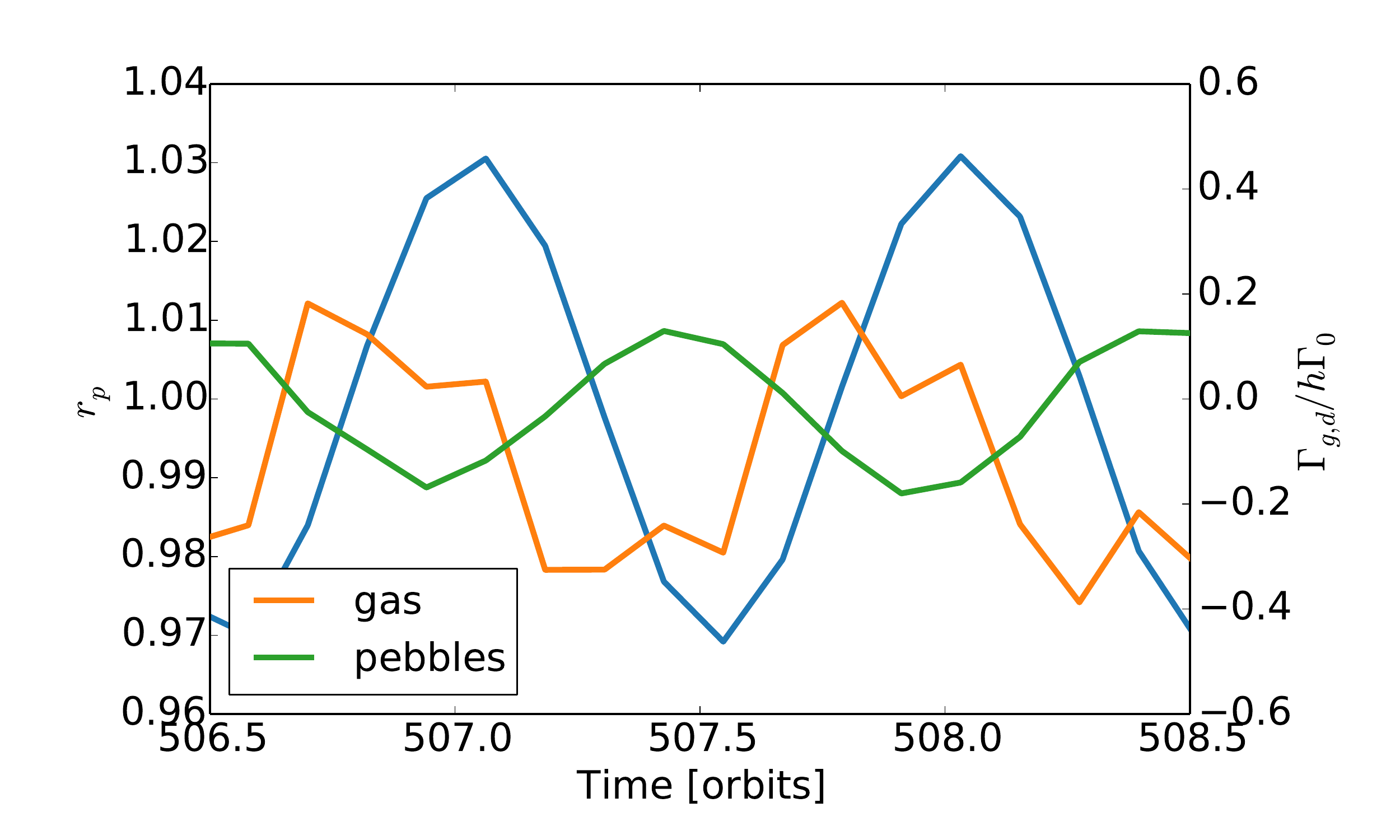}
\caption{For our reference run with $\dot M=2\times 10^{-8} M_\odot/$yr and a luminous $1$ $M_\oplus$ planet, time evolution of the gas and pebble torques over a few planetary orbits. The blue line shows the time evolution of the planet orbital distance.}
\label{fig:torques_orbit}
\end{figure}

Comparing the left and right panels of Fig. \ref{fig:torques}, we immediately see that a main effect of heat release in the disc is to almost cancel the torque induced by the solid component.  The relative pebble surface density perturbation at different times is shown  in the upper panel of Fig.\ref{fig:2d}. This figure suggests that  torques induced by the solid component in the disc are suppressed as a result of eccentricity growth. At early times, the underdense region located behind the planet and  responsible for the positive torque acting of the planet (Regaly 2020) is clearly visible. At later times, when the eccentricity has reached a saturated value, however, radial excursion of the planet in the disc causes the pebble distribution to adopt a complex shape over one orbital period.  Examining the evolution of the pebble torque over a few orbits in Fig.  \ref{fig:torques_orbit} reveals that it is not only positive (resp. negative) at pericentre (resp. apocentre), but also that it tends to be null on average.  Also, we note that a  test simulation performed with $f_{red}=0$ also resulted in a similar outcome, which confirms that the cancellation of the solid torque is not due to the removal of solids.

The gas torque exhibits similar oscillations over one planetary orbit and which can be explained by looking at the gas distribution in the lower panel of Fig.  \ref{fig:2d}.  We see that once the eccentricity has reached a saturated value,  there is a clear trend for the effect of heat release in the disc to manifest itself as underdense structures forming  behind (resp. in front of ) the planet at pericentre (resp. apocentre) and that exert a positive (resp. negative) torque on the planet.   In Fig. \ref{fig:torques_orbit}, the slight shift between the two curves corresponding to the evolution of $\Gamma_g$ and to that of  the planet orbital distance $r_p$ is due to   the time needed for the planet to create the underdense structures at apocentre/pericentre.  At this stage, eccentricity saturates because  asymetries in the immediate embryo vicinity, namely inside the Hill sphere,  weaken  as a result of the planet radial excursion (Chrenko et al. 2017). \\

\begin{figure}
\centering
\includegraphics[width=\columnwidth]{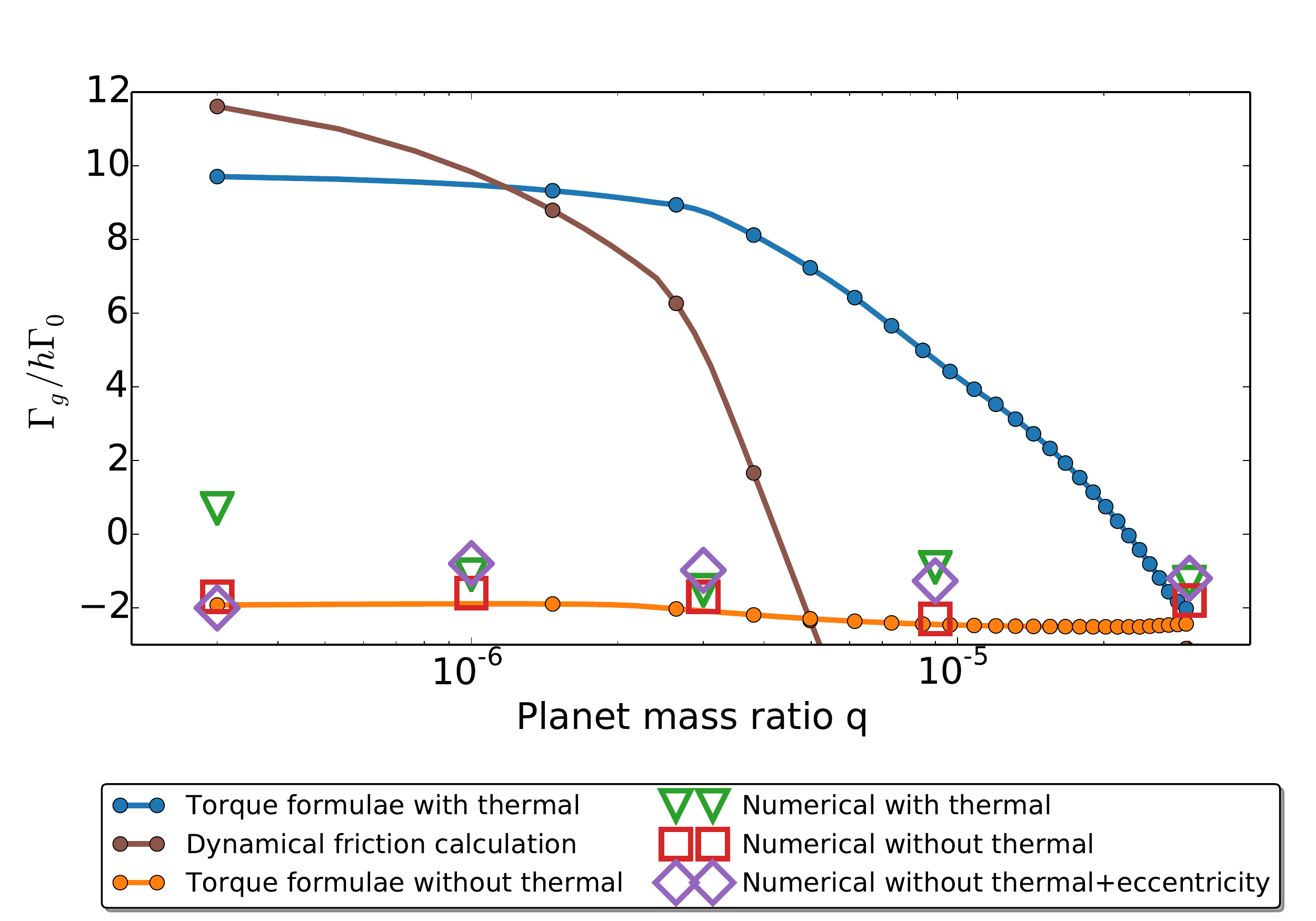}
\caption{For the fiducial case with $\dot M=2\times 10^{-8}$$M_\odot/$yr, gas torques as a function of planet-to-star mass ratio in various cases. Torque formulae correspond to those of Jimenez \& Masset (2017)  for the Lindblad+corotation torques and to that of Masset (2017) for the thermal torque.}
\label{fig:torques_vs_q}
\end{figure}

This suggests that we expect thermal torques to decrease as the eccentricity increases, and this seems to be confirmed by looking at the time evolution of the gas torques in the luminous planet case (bottom left panel of Fig. \ref{fig:torques}). From the above discussion in Sect. \ref{sec:a}, it seems indeed unlikely that the decrease in gas torques that is observed is caused by a reduction in the corotation torque. To further enlighten the role of eccentricity on the cut-off of thermal torques, we show in Fig.  \ref{fig:torques_vs_q} the torque as a function of $q$ in various cases. We first compare the gas torques at steady-state obtained in the luminous and non-luminous planet calculations with the results of an additional set of simulations of non-luminous planets evolving on fixed eccentric orbits. The planet eccentricity in these runs is set to the saturated value obtained in the luminous planet case, and therefore depends on the  value for $q$. The fact that this procedure results in torque values that agree fairly well with the torques obtained in situations where the planet releases heat in the disc suggests that thermal torques converge to zero in this latter case. This is true for all planet masses except for $q=3\times 10^{-7}$,  but we notice that we have not reached complete convergence for this run due to  high computational cost.

In Fig.  \ref{fig:torques_vs_q}, we also compare our derived torque values with existing analytical formula for the torque. These include the  prescriptions of Jimenez \& Masset (2017) for the Lindblad and corotation torques contributions, plus the expression of Velasco Romero \& Masset (2020) for the thermal torque. It takes into account the decay of the thermal torque for masses above the critical mass,

$$
M_{\rm crit}=\chi c_s/G,
$$

 which is estimated here to $M_{\rm crit}\approx 0.15$ $M_\oplus$ ($q\approx 5\times 10^{-7}$) , and is given by:

\begin{equation}
\Gamma_t=\frac{4M_{\rm crit}}{m_p+4M_{\rm crit}}\Gamma_{\rm heat}+\frac{2M_{\rm crit}}{m_p+2M_{\rm crit}}\Gamma_{\rm cold}
\label{eq:total_torque}
\end{equation}

with:

\begin{equation}
\gamma\frac{\Gamma_{\rm cold}}{\Gamma_0}=-1.61(\gamma-1)\frac{x_p}{\lambda_c}
\end{equation}

and

\begin{equation}
\gamma\frac{\Gamma_{\rm heat}}{\Gamma_0}=1.61(\gamma-1)\frac{x_p}{\lambda_c}\frac{L}{L_c}
\label{eq:gamma_heat}
\end{equation}

 where the value for $L/L_c$ in Eq. \ref{eq:gamma_heat} can be deduced from Fig. \ref{fig:lumin},  and  where $x_p$ is the distance from the planet to its corotation which  is estimated to  $x_p\approx 0.5 \lambda_c$  for the disc parameters adopted here. Although the analytical torque formulae of Masset (2017) have been derived in the limit $x_p/\lambda_c \rightarrow 0$, 
it has been shown that a corotation  offset with $x_p/\lambda_c\approx 0.5$ still yield to torques comparable to those given by linear theory (see Fig. 3 in Chamelta \& Masset 2021).  In the expressions above, $\Gamma_t$ is the thermal torque and $\Gamma_{\rm cold}$ (resp. $\Gamma_{\rm heat}$) stands for the cold thermal (resp. heating) torque.  Although the torque values in the non-luminous planet case are consistent with those predicted by the formula of Jimenez \& Masset (2017), we see that there are significant discrepancies between the numerical and analytical torques when thermal torques are included. \\

 It can not be excluded that the  differences that are observed are due to a lack or resolution, or occur because torque formulae have been derived using a 3 dimensional disc model whereas in this work we employ a 2 dimensional setup. However, looking back to the bottom left panel of Fig. \ref{fig:torques}, we can see that at early times when the thermal torques are established, values for the gas torques are consistent  with analytical formulae estimations. This indicates that our employed numerical resolution, together with considering a two dimensional model, are probably enough to obtain a reliable value of thermal torques, at least in the limit of small eccentricities.  An alternative origin of the discrepancy could be  that using Eq. \ref{eq:total_torque} becomes no longer valid when the planet acquires significant eccentricity.  As the eccentricity grows, the plume size whose an estimation is given by  $\lambda\approx \chi/\gamma V_p$ (Masset \& Velasco Romero 2017), where $V_p$ is the planet velocity relative to the gas,  can indeed become  equivalent to the distance to corotation $x_p$ and one has to resort to a dynamical friction calculation to evaluate the force acting on the perturber. Given that $V_p\approx a_p e_p \Omega$, we find $\lambda\sim x_p$ as soon as $e_p\gtrsim 0.01$.  This condition should also approximately coincide with a response time of the drag force (Masset \& Velasco Romero 2017) $\tau_{DF}=\chi/V_p^2$ shorter than the shear time scale $\Omega^{-1}$. In that case, we find  that for $e_p\sim 0.003$, the effect of the shear becomes negligible  and the torque can be obtained by a dynamical friction calculation. In this  context, which has been referred to as the headwind-dominated regime (Eklund \& Masset 2017), the net thermal torque exerted on a subsonic planet  has been derived by  Velasco Romero \& Masset (2020) and is given by: 

\begin{equation}
\Gamma_t=\Gamma_{adi}+\frac{4M_{\rm crit}}{m_p+4M_{\rm crit}}\Gamma_{\rm heat}+\frac{2M_{\rm crit}}{m_p+2M_{\rm crit}}\Gamma_{\rm cold}
\label{eq:total_torque_df}
\end{equation}

with:

\begin{equation}
\Gamma_{\rm heat}=\sqrt{2\pi}\frac{\gamma-1}{\gamma}\frac{L}{L_c}\Gamma_0,
\end{equation}

\begin{equation}
\Gamma_{\rm cold}=-\sqrt{2\pi}\frac{\gamma-1}{\gamma}\Gamma_0
\end{equation}

and:

\begin{equation}
\Gamma_{\rm adi}=-2\sqrt{2\pi}\frac{\gamma-1}{\gamma}\frac{\cal M}{3}\Gamma_0
\end{equation}

where $\cal M$ is the Mach number. Similarly to Eq. \ref{eq:total_torque}, the decay of thermal torques past the critical mass $M_c$ is taken into account in Eq.  \ref{eq:total_torque_df}. We also notice that since $e_p\lesssim h$  when the eccentricity saturates, $\cal M \lesssim $ 1 such that the assumption of considering the subsonic regime is justified. In the supersonic regime, the net torque is expected to decrease as   $V_p^2$,  but this occurs for values of the Mach number $\cal M > $1 (Velasco Romero \& Masset 2019).  The net torques obtained in the dynamical friction approximation is represented by the brown line in Fig. \ref{fig:torques_vs_q}.  Again, these clearly overestimate the thermal torques that we obtained in the simulations, which further supports an additional cut-off of thermal torques as the planet acquires significant eccentricity.

\subsection{A prescription for the thermal torque cut-off}

\begin{figure}
\centering
\includegraphics[width=\columnwidth]{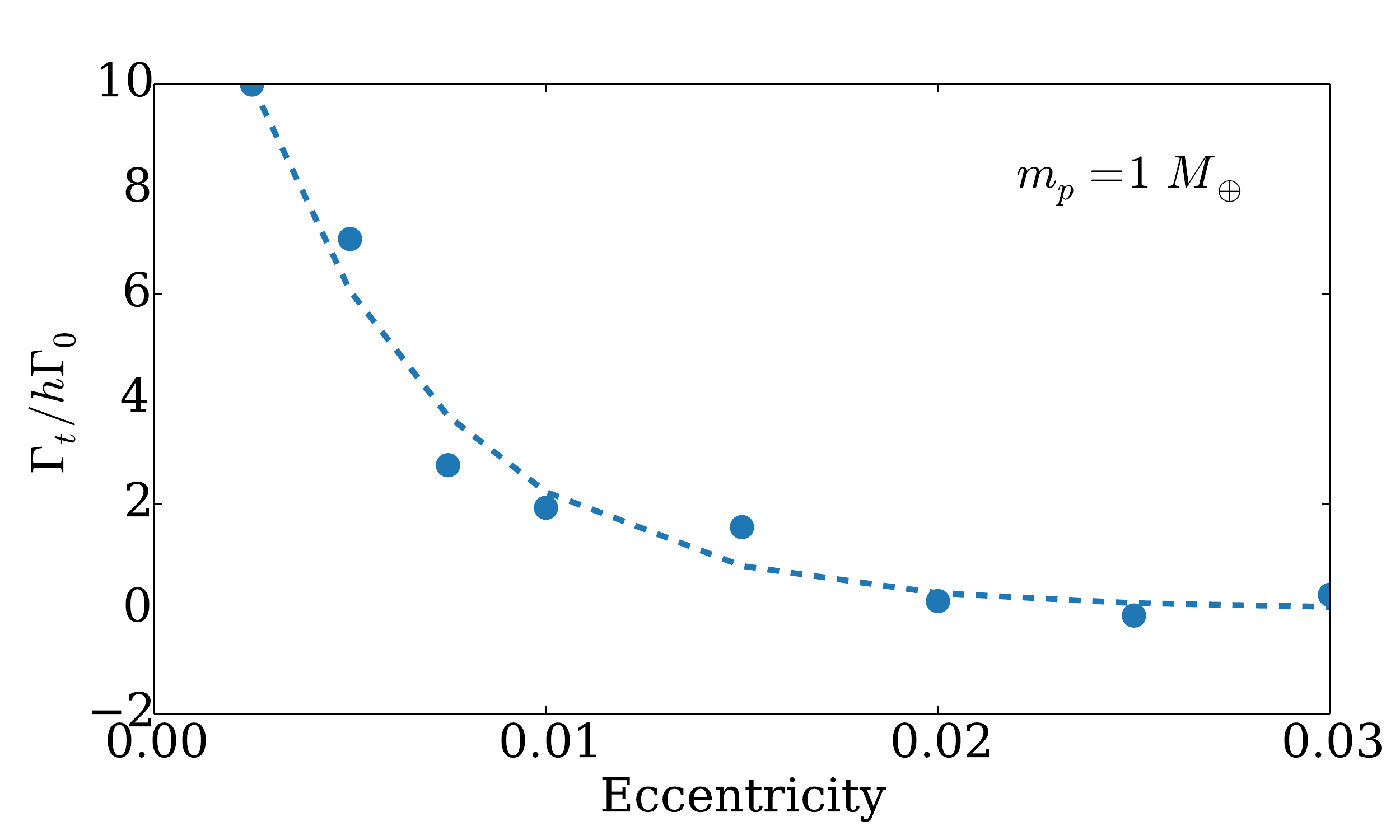}
\caption{ Thermal torque as a function of eccentricity for a $1$ $M_\oplus$ embryo measured in our simulations (filled circles). The dashed line corresponds to a fit of the form given by Eq.\ref{eq:fit} with $e_f=0.07\lambda_c/H$.}
\label{fig:fit}
\end{figure}

As mentionned above, these results can be interpreted as a consequence of the decay of thermal torques with increasing eccentricity.  The aim of this section is to provide an empirical fit to our data, assuming that the thermal torques $\Gamma_t$ decrease exponentially with eccentricity as:

\begin{equation}
\Gamma_t=\Gamma_{t,e_p=0}\exp\left(-\frac{e_p}{e_f}\right)
\label{eq:fit}
\end{equation}

where $e_f$ is an e-folding eccentricity which is expected to be a fraction of $\lambda_c/H$ (Paardekooper et al. 2022) . To determine $e_f$, we consider a $1$ $M_\oplus$ embryo evolving on fixed eccentric orbits with $e_p\in[0,0.03]$. For a non-luminous planet, we recall that the total torque is given as the sum of the Lindblad plus the attenuated corotation torque due to the planet having finite eccentricity (Fendyke \& Nelson 2014), while for a luminous planet, there is also a contribution from the thermal torque. Hence, thermal torques can be simply obtained by taking the difference between the torques in these two sets of simulations.  Results of this procedure are shown in Fig. \ref{fig:fit} which displays the time averaged thermal torques as a function of eccentricity, and where the superimposed line in the plot corresponds to Eq. \ref{eq:fit} using the best fitting value for $e_f$ that we obtain. It is given by:

\begin{equation}
e_f\approx 0.07\frac{\lambda_c}{H}
\label{eq:ef}
\end{equation}

\begin{figure}
\centering
\includegraphics[width=\columnwidth]{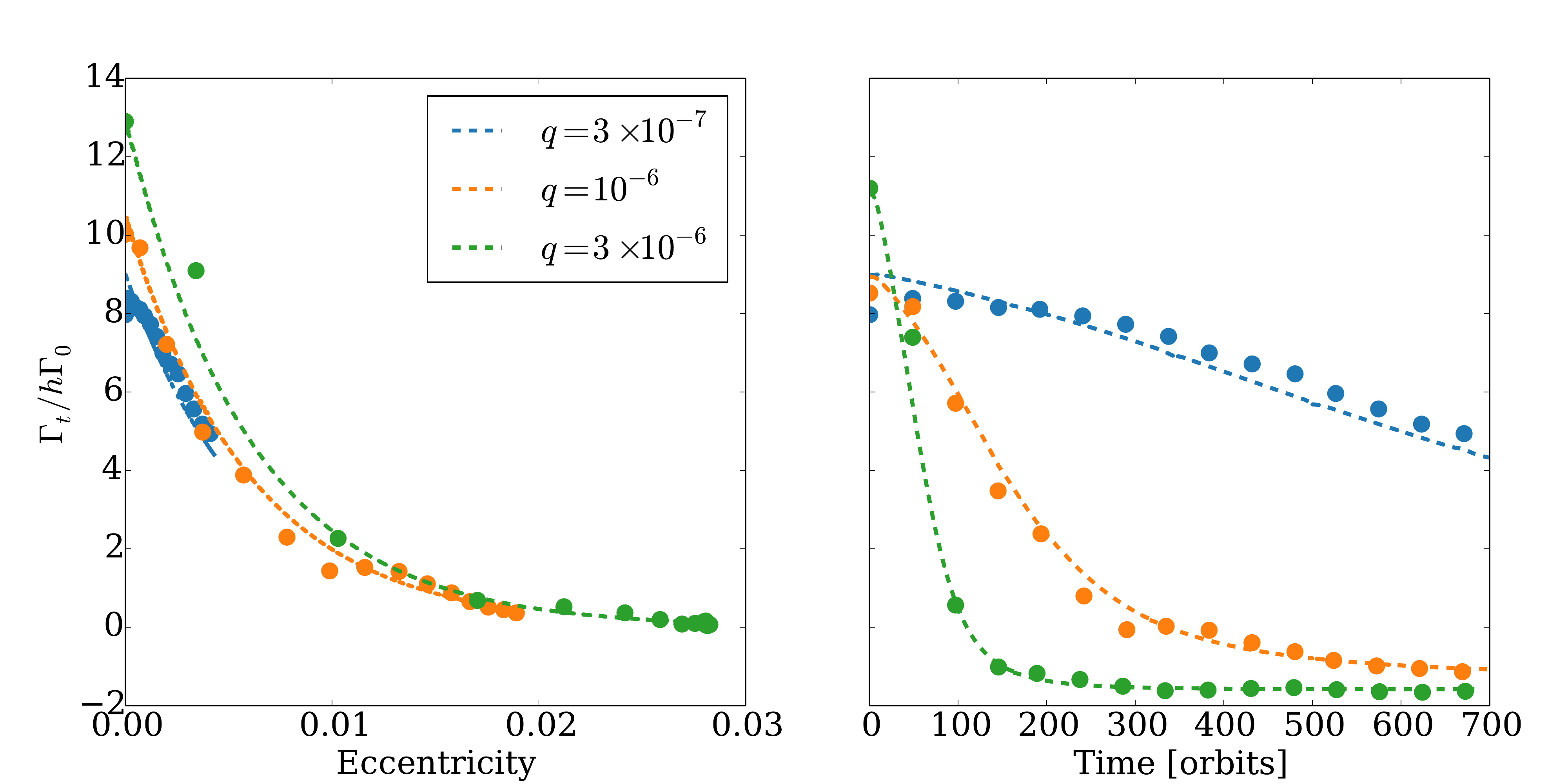}
\caption{Left panel: thermal torque as a function of eccentricity (filled circles). The dashed line corresponds to a fit of the form given by Eq.\ref{eq:fit} with $e_f=0.07\lambda_c/H$. Right panel:  thermal torque as a function of time as measured in our simulations (filled circles). The dashed line corresponds to a fit of the form given by Eq. \ref{eq:fit_time}.}
\label{fig:curve_fit}
\end{figure}

Going back to Fig. \ref{fig:torques_vs_q}, we see that for our fiducial model, the torques experienced by non-luminous planets evolving on fixed circular and eccentric orbits are in fact comparable. We remind the reader that in the eccentric case,  the eccentricity is set to the value reached by the embryo when thermal torques are included.  This motivates us to check whether Eqs.  \ref{eq:fit} and \ref{eq:ef} provide reasonable fits of  thermal torques, assuming that these can be determined by taking the difference between the total torques obtained in the luminous (bottom left panel in Fig. \ref{fig:torques}) and non-luminous  (bottom right  panel in Fig. \ref{fig:torques}) cases.  The left panel of Fig. \ref{fig:curve_fit} shows that our fitting formula agrees reasonably with the data. The right panel compares the thermal torques with the function:

\begin{equation}
\Gamma_t(t)=\Gamma_{t,e_p=0}\exp\left(-\frac{e_p(t)}{0.07(\lambda_c/H)}\right)
\label{eq:fit_time}
\end{equation}

where $e_p(t)$ is plotted as a function of time in Fig. \ref{fig:ae}. Again, this formula provides a good fit of the data.  Here, it is worthwhile to note  that care must be taken when employing the previous prescription for the decay of thermal torques with eccentricity in the context of  a different source of luminosity (e.g. planetesimal accretion, accretion onto stellar-mass black holes in AGNs...), since in this work the accretion efficiency arising from pebble accretion is itself a function of eccentricity. We emphasize, however, that for a $1$ $M_\oplus$ planet, the pebble accretion efficiency varies from $\epsilon \approx 0.085$ for $e_p=0$, to  $\epsilon \approx 0.1$ for $e_p=0.03$ such that this effect is probably marginal.  

\begin{figure}
\centering
\includegraphics[width=\columnwidth]{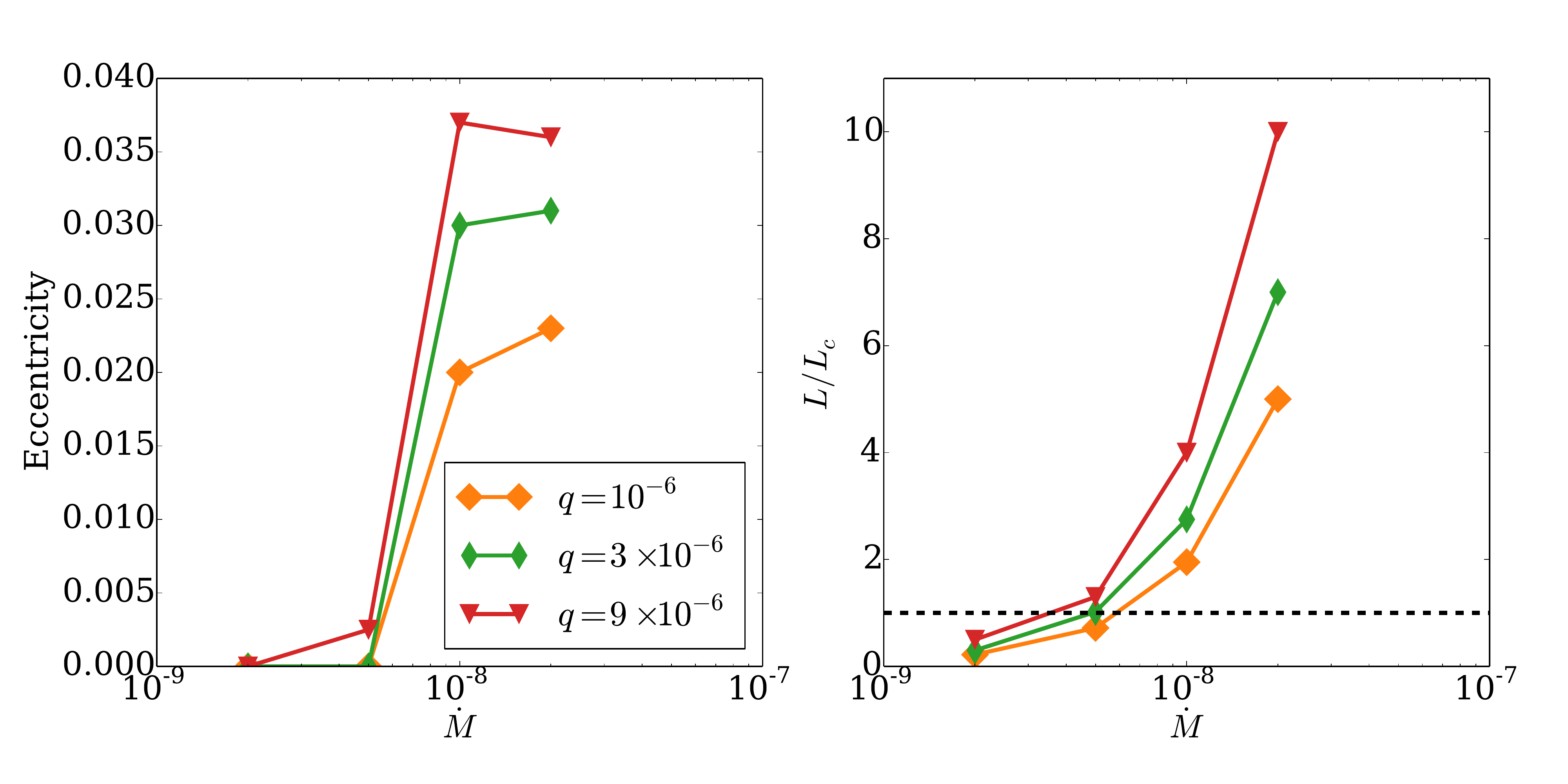}
\caption{Left: Eccentricity at saturation phase as a function of accretion rate for various values of the planet-to-star mass ratio. Right: same but for the luminosity. Growth of eccentricity is not expected for luminosities $L<L_c$ (dashed line).}
\label{fig:le_mdot}
\end{figure}

\begin{figure*}
\centering
\includegraphics[width=0.32\textwidth]{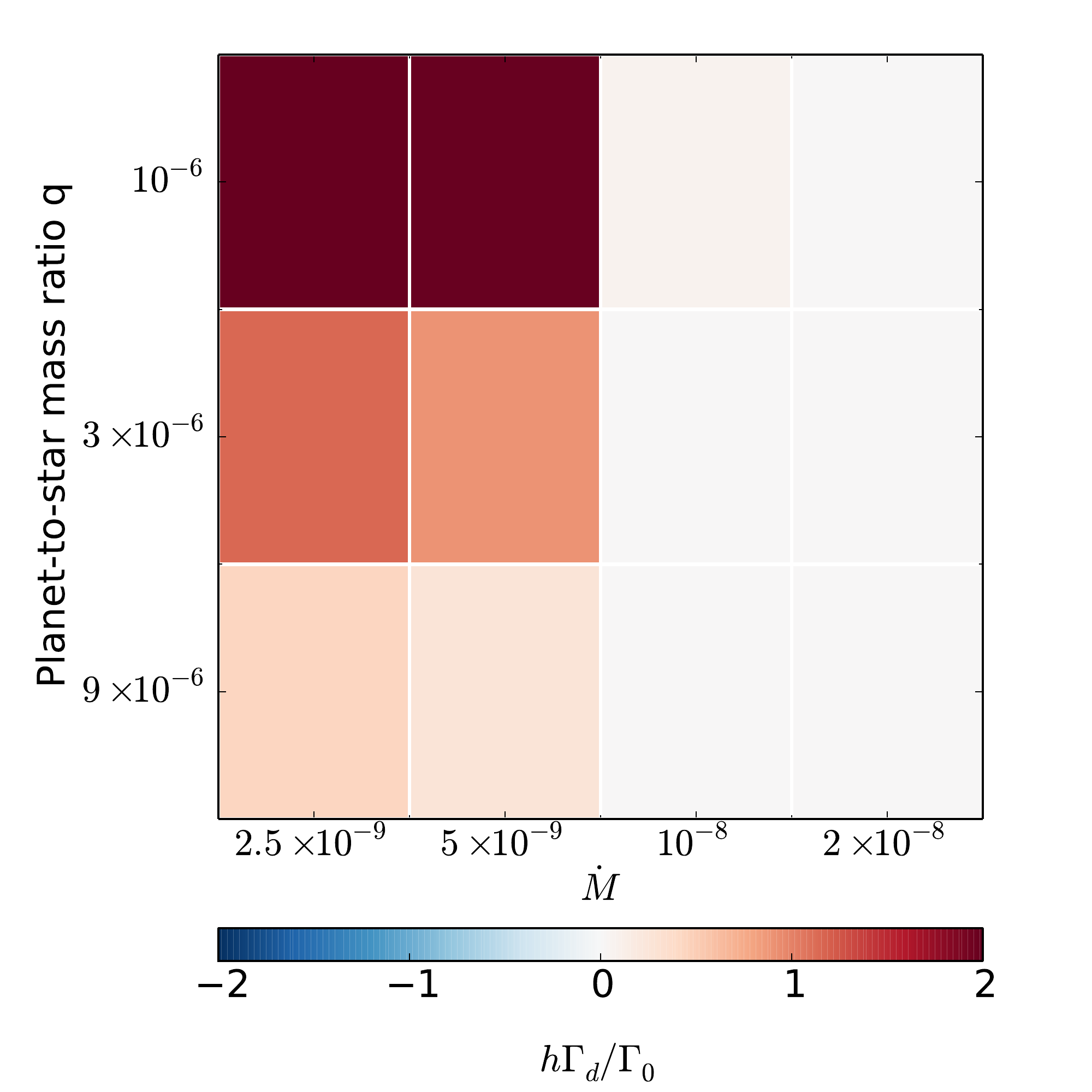}
\includegraphics[width=0.32\textwidth]{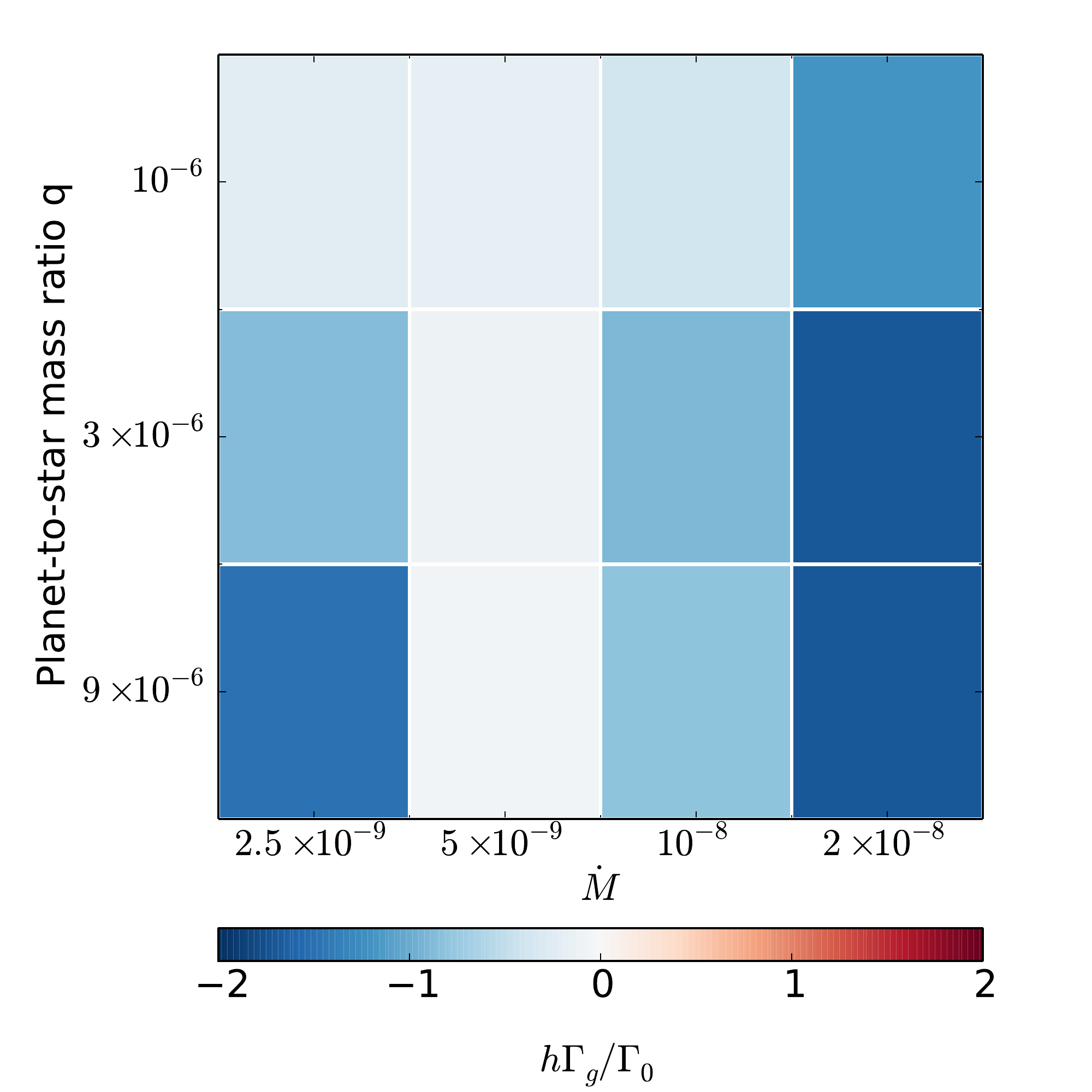}
\includegraphics[width=0.32\textwidth]{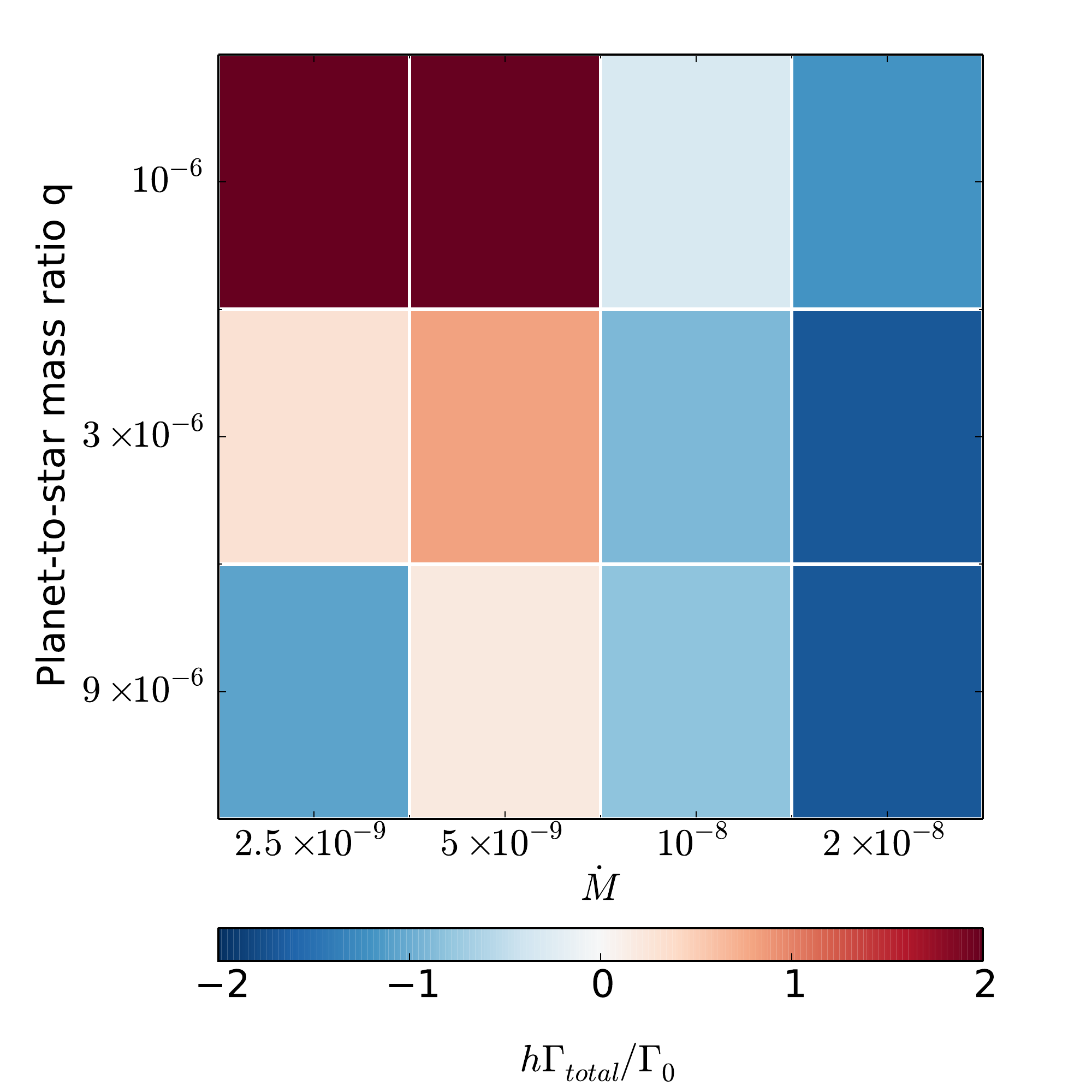}
\caption{From left to right, torque maps showing the strength of dust $\Gamma_d$, gas $\Gamma_g$, and total torques $\Gamma_{total}$ as a function of disc accretion rate $\dot M$ and planet-to-star mass ratio. }
\label{fig:maps}
\end{figure*}

\section{Dynamics in evolving protoplanetary discs}

In this section, we test the influence of varying the accretion rate due to the wind $\dot M_w$ for which we considered values $\dot M_w\in [2\times 10^{-9},2\times 10^{-8}]$ $M_\odot/$yr. According to Eq. \ref{eq:mdot}, this would span disk ages from $t_{\rm disk}=5\times 10^5$ yr (for $\dot M=2\times 10^{-8}$ $M_\odot/$yr) to $t_{\rm disk}=3$ Myr (for $\dot M=2\times 10^{-9}$ $M_\odot/$yr). The pebble flux is calculated from Eq. \ref{eq:mdot_peb} and continuously decreases with $\dot M$ down to $\dot M_{\rm peb}\sim 70$ $M_\oplus$/Myr for $\dot M=2\times 10^{-9}$ $M_\odot /$yr. We show in Fig. \ref{fig:le_mdot} the equilibrium eccentricity and luminosity of the embryo as a function of $\dot M$. Eccentricity growth is not observed  at luminosities $L\lesssim L_c$, consistently with previous work (Masset 2017; Velasco Romero et al. 2021). From the results described in Sect. \ref{sec:fiducial}, the  torque exerted by the solid component is not expected to cancel for such sub-critical luminosities. The relative importance between dust and gas torques is highlighted in the left panel of Fig. \ref{fig:maps} which summarizes the results of our numerical exploration. From left to right, we plot the dust, gas and total torques respectively  a a function  of $\dot M$ and $q$.  In line with expectation, dust torques tend to cancel for $\dot M \gtrsim 10^{-8}M_\odot/$yr, where eccentricity growth occurs at all planetary masses. 

As the disc evolves and the pebble flux decreases, however, the luminosity becomes  smaller and smaller and eventually decreases below  the critical luminosity $L_c$. This prevents the embryo eccentricity to grow, such that the torque exerted by the solid component becomes non-zero again. This is particularly true for embryo mass $\lesssim 1$ $M_\oplus$ for which positive dust torques tend to overcome gas torques for accretion rates $\dot M \lesssim 5\times 10^{-9}$ $M_\odot/$yr, as revealed by the total torque map in the right panel of Fig. \ref{fig:maps}. Interestingly, this imply that low-mass embryos may undergo pebble-driven outward migration in the late stages of the disc evolution.

\section{Conclusions}

In this paper we studied  the orbital evolution of low-mass embryos with mass $m_p\in[0.1,3]$ $M_\oplus$ in protoplanetary discs in the context of a pebble accretion scenario. As an embryo grows by accreting pebbles,  heat release  in the vicinity of the planet gives rise to a thermal force whose primary effect is to lead to planet eccentricity growth when the accretion luminosity is high enough. Using two-dimensional hydrodynamical simulations, we examined how the evolution outcome depends on disk age and evaluated the relative importance of thermal torques as the disc evolves and as the gas and pebble mass fluxes drop.  We considered a disc model where accretion occurs mainly through a disc wind generated by the Hall effect, and which we modelled by applying a synthetic torque onto the disc. We also considered the feedback between the embryo luminosity and eccentricity by using the prescriptions for the  pebble accretion efficiencies  of Liu \& Ormel (2018) and Ormel \& Liu (2018).\\
At disc evolution time $t_{\rm disk}=5\times 10^5$ yr, which corresponds to an accretion rate of $\dot M= 2\times 10^{-8}$ $M_\odot/$yr, typical initial evolution of pebble-accreting cores involves outward migration plus eccentricity growth due to significant heating torques. However, Earth-mass embryos are found to ultimately migrate inward due to a thermal torque cut-off associated with the planet radial excursion in the disc. We indeed found that the amplitude of thermal torques decrease exponentially with increasing eccentricity, and we provided a fitting formula for the thermal torque attenuation as a function of eccentricity which might be included in N-body models of planet formation that include thermal torques effects.   Our results also imply that the cut-off of  thermal torques can be twofold: i)  These can decay when the mass of the perturber is larger than the critical mass $M_c=\chi c_s/G$ (Velasco-Romero \& Masset 2020), because in this situation not all of the energy released by the planet is converted in gas heating outside of the Bondi sphere.  In that case,  thermal torques are expected to decay as $m_p^{-1}$ past the critical mass; ii) thermal forces can also weaken due to the finite value of the planet eccentricity, in which case the decay of thermal torques with eccentricity is given by Eqs. \ref{eq:fit} and \ref{eq:ef}, independently of the planet mass.

   A related effect of the eccentricity growth is that the  torque exerted by the solid component cancels  on average, whereas it tends to be positive for a planet evolving on a fixed circular orbit.\\
As the disc evolves and the radial flux of pebbles drops, however, the accretion luminosity becomes too small to make the core eccentricity grow, which allows again for a non-zero (positive) torque induced by the pebbles. Interestingly, we find that for accretion rates $\dot M \lesssim 5\times 10^{-9}$ $ M_\odot/$yr, the pebble torque exerted on cores with mass $m_p\lesssim$ $1M_\oplus$ can even overcome the gas torque, resulting in outward migration.\\
One main limitation here is that we only  performed 2D simulations and future work should adopt a more realistic 3D setup.  It has been indeed shown that the streamlines outflowing from the Hill sphere may be significantly distorted in 3D (Chrenko \& Lambrechts 2019) such that we expect significant differences in the heating torque magnitude between 2D and 3D. Given that in this work we considered temperature-dependent opacities, we also expect oscillations in the heating torque to arise when moving to 3D (Chrenko \& Lambrechts 2019).

\section*{Acknowledgments}
Computer time for this study was provided by the computing facilities MCIA (M\'esocentre de Calcul Intensif Aquitain) of the Universite de Bordeaux and by HPC resources of Cines under the allocation A0110406957 made by GENCI (Grand Equipement National de Calcul Intensif). 
\section*{Data Availability}

The data underlying this article will be shared on reasonable request to the corresponding author.

\appendix

\section{Pebble accretion efficiency for eccentric planets}

In this section, we give the prescriptions of Liu \& Ormel (2018) and Ormel \& Liu (2018) for the pebble accretion efficiency of eccentric planets and that we used in our simulations. The pebble accretion efficiency on the 2D settling regime is given by:

\begin{equation}
\epsilon_{2D,\rm set}=0.32\sqrt{\frac{q}{\st \;\eta^2}\frac{\Delta V}{v_k}}f_{\rm set}
\label{eq:a1}
\end{equation}
where $v_k$ is the keplerian velocity and $\Delta V$  the pebble-embryo relative velocity which is given by:

\begin{equation}
\Delta V=\max(V_{\rm circ}, V_{\rm ecc})
\end{equation}

with:

\begin{equation}
V_{\rm ecc}=0.76e_p v_k
\end{equation}

and:

\begin{equation}
V_{\rm circ}=\left[1+5.7\left(\frac{q\st}{\eta^3}\right)\right]^{-1}+0.52(q\st)^{1/3}v_k
\end{equation}

Moreover, in Eq. \ref{eq:a1}, the transition function is given by:
\begin{equation}
f_{\rm set}=\exp\left[-0.5\frac{\Delta V}{V_\ast^2}\right]\times\frac{V_\ast}{\sqrt{V_\ast^2+0.33\sigma_{pz}^2}}
\end{equation}

where $V_\ast$ is the transition velocity:
\begin{equation}
V_\ast=\left(\frac{q}{\st}\right)^{1/3}v_h
\end{equation}

and $\sigma_{pz}$ is the vertical turbulent velocity (Youdin \& Lithwick 2007):

\begin{equation}
\sigma_{pz}=\frac{\alpha}{1+\st}\left(1+\frac{\st}{1+\st}\right)^{-1/2}hv_k
\end{equation}

The pebble accretion in the 3D settling regime is given by:

\begin{equation}
\epsilon_{3D,\rm set}=0.39\frac{q}{\eta h_d}f_{\rm set}^2
\end{equation}

with $h_d$ the pebble disc aspect ratio  (Youdin \& Lithwick 2007):

\begin{equation}
h_d=\sqrt{\frac{\alpha}{\alpha+\st}}\left(1+\frac{\st}{1+\st}\right)^{-1/2}h
\end{equation}

 In this regime, the accretion radius $R_{acc}$  is given by:

\begin{equation}
R_{acc}=\sqrt{\frac{Gm_p t_s}{\Delta V}},
\end{equation}

Finally, the accretion efficiency in the 2D and 3D ballistic regimes are respectively given by:
\begin{equation}
\epsilon_{2D,\rm bal}=\frac{R_p}{2\pi \eta R \st}\sqrt{\frac{2qR}{R_p}\left(\frac{\Delta V}{v_k}\right)^2}(1-f_{\rm set})
\end{equation}

and:
\begin{equation}
\epsilon_{3D,\rm bal}=\frac{1}{4\sqrt{2\pi} \eta h_d \st}\left(2d\frac{v_k}{\Delta V}\frac{R_p}{R}+\frac{R_p^2}{R^2}\frac{\Delta V}{v_k}\right)(1-f_{\rm set}^2)
\end{equation}

 In the ballistic regime, $R_{acc}$ reads:

\begin{equation}
R_{acc}=R_p\sqrt{\left(\frac{V_{esc}}{\Delta V}\right)^2+1}
\end{equation}

where $V_{esc}=\sqrt{\frac{2Gm_p}{R_p}}$ is the escape velocity.\\

The expression that we employ to calculate the pebble accretion efficiency in our calculations is given by:

\begin{equation}
\epsilon=\frac{f_{\rm set}}{\sqrt{\epsilon_{2D,\rm set}^{-2}+\epsilon_{3D,\rm set}^{-2}}}+\frac{1-f_{\rm set}}{\sqrt{\epsilon_{2D,\rm bal}^{-2}+\epsilon_{3D,\rm bal}^{-2}}}
\end{equation}

\end{document}